\documentclass[preprint2]{proto}
\usepackage{times}
\usepackage{graphicx}
\usepackage{wasysym}
\usepackage{float}
\usepackage{dblfloatfix}
\usepackage{pifont}
\newcommand{\cmark}{\ding{51}}%
\newcommand{\xmark}{\ding{55}}%

\usepackage{soul}
\usepackage[dvipsnames]{xcolor}

\graphicspath{{./}{Figures/}}

\begin{document}

\title{\textbf{\LARGE COMETS AND METEOR SHOWERS}}

\author {\textbf{\large Quanzhi Ye}}
\affil{\small\em Department of Astronomy, University of Maryland, College Park, Maryland 20742, USA}
\affil{\small\em Center for Space Physics, Boston University, 725 Commonwealth Ave, Boston, Massachusetts 02215, USA}

\author {\textbf{\large Peter Jenniskens}}
\affil{\small\em SETI Institute, 339 Bernardo Avenue, Mountain View, California 94043, USA}
\affil{\small\em NASA Ames Research Center, Moffett Field, California 94035, USA}




\begin{abstract}

\begin{list}{ } {\rightmargin 1in}
\baselineskip = 11pt
\parindent=1pc
{\small 
Earth occasionally crosses the debris streams produced by comets and other active bodies in our solar system. These manifest meteor showers that provide an opportunity to explore these bodies without a need to visit them in-situ. Observations of meteor showers provide unique insights into the physical and dynamical properties of their parent bodies, as well as into the compositions and the structure of near-surface dust. In this chapter, we discuss the development and current state of affairs of meteor science, with a focus on its role as a tool to study comets, and review the established parent body -- meteor shower linkages.
\\~\\~\\~}
\end{list}
\end{abstract}  

\section{\textbf{INTRODUCTION}}
\label{sec:intro}

Comets account for most of the mass of the interplanetary dust cloud and most meteoroids colliding with Earth \citep{Nesvorny2010, Jenniskens2015}. They actively release gas into interplanetary space, and with it solid particles described alternatively as interplanetary dust, meteoroids, or cometary fragments if they are larger than 1~m in diameter. Lighter materials, such as gas and sub-micron-sized dust (also known as $\beta$-meteoroids), are carried by the solar wind directly outward from the Sun into interstellar space. Most heavier materials, such as millimeter-size or larger meteoroids, do not escape the gravity of the Sun and planets, but are still influenced by the radiation forces. A small fraction ends up colliding with Earth's atmosphere. The phenomenon created by the entry of a meteoroid into Earth's atmosphere is known as a {\it meteor}. What survives to the ground is called a {\it meteorite}. 

\begin{figure*}[ht!]
\begin{center}
\includegraphics[width=15cm]{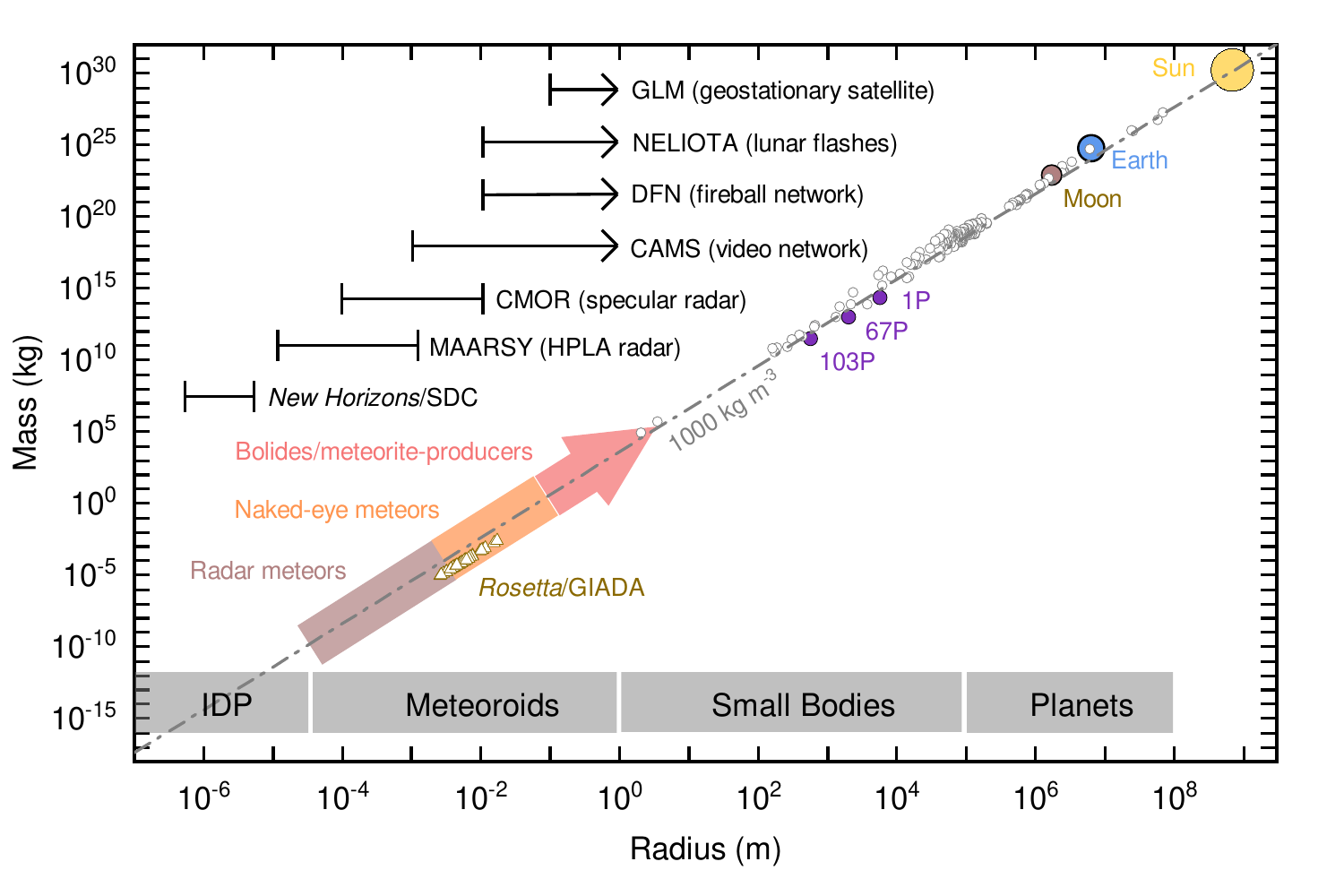}
\caption{A diagram of mass and sizes of various bodies in the solar system, labeled with the 2017 IAU definition of meteoroid and interplanetary dust particle (IDP). Also shown are the size regimes of radar, naked-eye (photographic/video) techniques, bolides/meteorite-producers, and examples of several observation programs: the Geostationary Lighting Mapper (GLM) as an example of orbit-based meteor surveillance, Near-Earth object Lunar Impacts and Optical TrAnsients (NELIOTA) for lunar flashes observations, the Desert Fireball Network (DFN) for fireball photography, the Cameras for All-Sky Meteor Surveillance (CAMS) for low-light video camera surveillance, the Canadian Meteor Orbital Radar (CMOR) for meteor orbit radars, the Middle Atmosphere Alomar Radar System (MAARSY) for High Power Large Aperture (HPLA) radar, and the Student Dust Counter (SDC) aboard New Horizons for space-based dust detector. For reviews of video and radar meteor programs, see \citet{Koten2019} and \citet{Kero2019}.}
\label{fig:def}
\end{center}
\end{figure*}

Historically, the terms ``(cometary/interplanetary) dust'' and ``meteoroid'' have been ambiguous and they are often used interchangeably. The comet community tends to use ``dust'', while the meteor community prefers ``meteoroid''. A clarification issued by the International Astronomical Union (IAU) \citep{Koschny2017} defined a meteoroid to be a solid natural object of a size roughly between $\sim30~\micron$ to 1~m, with interplanetary dust being the solid matter smaller than meteoroids, but acknowledged that the size distribution is continuous (Figure~\ref{fig:def}). In this chapter, we use these two terms interchangeably.

After one revolutions around the Sun, some meteoroids return early, others later, creating a stream. Those streams are sometimes seen in infrared emission as cometary dust trails (see the chapter by {\it Agarwal et al.}). Earth passing through a stream results in a large number of meteoroids entering the atmosphere from the same direction, forming a {\it meteor shower}. Due to perspective, observers see shower meteors radiate from a point on the celestial sphere (the {\it radiant}). The naming of meteor showers is governed by the IAU's Working Group on Meteor Shower Nomenclature. A meteor shower is usually named after the nearest bright star to the radiant at the time of peak activity. The month of the year is sometimes added to distinguish showers in the same constellation, and the word ``daytime'' is added to showers that are mostly active during daytime. A complete list of meteor showers is maintained by the IAU Meteor Data Centre (MDC) \citep{Jopek2017}. As of the writing, there are 112 ``established'' showers and 830 ``working list'' showers. Showers are assigned a unique name, a three-letter code, and a number which are all unique to each shower. For instance, the Leonids associated with comet 55P/Tempel--Tuttle have code LEO and number \#13, distinguishing it from the other 32 showers with radiants in the constellation of Leo throughout the year. 

Many physical parameters are derived from meteor and meteor shower research, including meteoroid orbits, measures of meteoroid deceleration, luminosity, electron density, fragmentation, elemental abundances, as well as stream dispersion, times of encounters, and flux density. An important observable is the Zenith Hourly Rate (ZHR), defined as the number of meteors seen by a single human observer with standard perception under perfect visual conditions (limiting stellar magnitude of +6.5 and shower radiant at zenith). ZHR is a proxy of the meteoroid stream's flux density, dependent on the magnitude distribution index of meteors (called ``population index'' in some meteor science literature, often represented by $r$ or $\chi$), which is related to the meteoroid size and mass distribution index, assuming meteoroid density and luminous efficiency are constant in a stream. The measured quantity of luminosity is most closely related to the kinetic energy of the meteoroid, not its mass or size \citep{Koschack1990}. The strongest annual meteor showers such as the Quadrantids, Perseids and Geminids have a ZHR around 100 (a volume density of mm-sized meteoroids around $10^{-12}$--$10^{-11}~\mathrm{m^{-3}}$), while the strongest meteor outbursts in the past two centuries had ZHR over $\sim$10,000 ($10^{-10}$--$10^{-9}~\mathrm{m^{-3}}$). Meteor activity with ZHR over 1,000 is called a \textit{meteor storm}.

\begin{deluxetable}{ccccc}
\tablecaption{Established\tablenotemark{a} linkages between comets/asteroids and confirmed meteoroid streams.\label{tbl:shr}}
\tablehead{
\colhead{Object} & \colhead{Type\tablenotemark{b}} & \colhead{Meteor shower and IAU ID number} & \colhead{ZHR\tablenotemark{c}} & \colhead{Mechanism\tablenotemark{d}}
} 
\startdata
1P/Halley & HTC & $\eta$-Aquariids (\#31), Orionids (\#8) & 60, 35 & S \\
2P/Encke\tablenotemark{e} & ETC & Daytime $\beta$-Taurids (\#173), N/S Taurids (\#17/\#2) & 10, 5 & S, BU? \\
3D/Biela & JFC & Andromedids (\#18) & $<2$\tablenotemark{f} & BU \\
7P/Pons--Winnecke & JFC & June Bootids (\#170) & $<2$\tablenotemark{f} & S \\
8P/Tuttle & JFC & Ursids (\#15) & 10\tablenotemark{f} & S \\
15P/Finlay & JFC & Arids (\#1130) & $<2$\tablenotemark{f} & S \\
21P/Giacobini--Zinner & JFC & October Draconids (\#9) & $<2$\tablenotemark{f} & S \\
26P/Grigg-Skjellerup & JFC & $\pi$-Puppids (\#137) & $<2$\tablenotemark{f} & S \\
55P/Tempel--Tuttle & HTC & Leonids (\#13) & 15\tablenotemark{f} & S \\
73P/Schwassmann--Wachmann 3 & JFC & $\tau$-Herculids (\#61) & $<2$\tablenotemark{f} & S? BU? \\
96P/Machholz 1\tablenotemark{g} & JFC & Daytime Arietids (\#171), N/S $\delta$-Aquariids (\#26/\#5) & 60, 30 & S + BU? \\
109P/Swift--Tuttle & HTC & Perseids (\#7) & 120 & S \\
169P/NEAT \& 2017 MB$_1$ & JFC & $\alpha$-Capricornids (\#1) & 5 & BU? \\
209P/LINEAR & JFC & Camelopardalids (\#451) & $<2$\tablenotemark{f} & S \\
289P/Blanpain & JFC & Phoenicids (\#254) & $<2$\tablenotemark{f} & S? + BU? \\
300P/Catalina & JFC & June $\epsilon$-Ophiuchids (\#459) & $<2$\tablenotemark{f} & S \\
C/1861 G1 (Thatcher) & LPC & April Lyrids (\#6) & 20\tablenotemark{f} & S \\
C/1911 N1 (Kiess) & LPC & Aurigids (\#206) & 5\tablenotemark{f} & S \\
C/1917 F1 (Mellish) & HTC\tablenotemark{h} & December Monocerotids (\#19) & 3 & S \\
C/1979 Y1 (Bradfield) & LPC & July Pegasids (\#175) & $<2$ & S \\
C/1983 H1 (IRAS–Araki-Alcock) & LPC & $\eta$-Lyrids (\#145) & $<2$ & S? \\
(3200) Phaethon & NEA & Geminids (\#4) & 180 & TBU? \\
(155140) 2005 UD & NEA & Daytime Sextantids (\#221) & 5 & TBU? \\
(196256) 2003 EH$_1$ & NEA & Quadrantids (\#10) & 130 & BU? \\
\enddata
\tablenotetext{a}{Here we only list associations that have been investigated/suggested by multiple studies from different research groups.}
\tablenotetext{b}{Acronyms: NEA -- near-Earth asteroid; ETC -- Encke-type comet; HTC -- Halley-type comet; JFC -- Jupiter-family comet; LPC -- Long-period comet.}
\tablenotetext{c}{Zenith Hourly Rate. See main text for details.}
\tablenotetext{d}{S -- sublimation; BU -- breakup; TBU -- thermal breakup.}
\tablenotetext{e}{Also known as the Taurid Complex. May include a couple of NEAs and other meteor showers \citep{Porubcan2006}.}
\tablenotetext{f}{Known to exhibit occasional outburst.}
\tablenotetext{g}{Also include the Marsden comet group (see \S~\ref{sec:obs:linkages}).}
\tablenotetext{h}{This comet has a orbital period of 143 yr, but is designated as a C/ comet.}
\end{deluxetable}

The history of meteor science has been discussed in great detail in, e.g. \citet[][\S~1]{Jenniskens2006}. Like comets, records of meteors go back to ancient times. The earliest confirmed sighting of a meteor shower is recorded in 687 BC in the Chinese chronicle {\it Zuo Zhuan}, when meteor showers were recognized as periods of higher meteor rates. The radiant phenomenon was widely recognized during the 1833 Leonids meteor storm, providing information on meteoroid orbits. The 1865 return of comet 55P/Tempel--Tuttle and the 1866 Leonid storm subsequently led to the recognition that meteor showers and comets are linked. Several other major meteor showers were also quickly associated with comets, especially comet 109P/Swift--Tuttle to the Perseids, comet C/1861 G1 (Thatcher) with the Lyrids, and 3D/Biela to the Andromedids. 

Since then, numerous efforts have been made to map meteor showers and to identify their parent bodies. Looking for radiants, observers in the 19th and early 20th century mostly relied on their naked eyes, a sky chart and a pencil to record meteor tracks. Photographic and radar techniques were introduced in the 1890s and 1920s, respectively. By the early 2000s, the number of meteoroid orbits was still relatively small. In 2002, the Canadian Meteor Orbit Radar (CMOR) started regular meteor observations in the northern hemisphere in radio wavelength, an effort that was later expanded to the southern hemisphere by the Southern Argentina Agile Meteor Radar (SAAMER). In 2007, early low-light video meteor triangulations were scaled up, first by the SonotaCo network in Japan, later expanded greatly by the European video MeteOr Network Database (EDMOND) and Cameras for All-Sky Meteor Surveillance (CAMS) video surveys. These techniques have ushered in a revolution in our understanding of meteor showers and an increasing number of associations with the parent bodies found by the Near-Earth Object (NEO) surveys. Table~\ref{tbl:shr} lists the well-established linkages between comets or active asteroids and meteoroid streams known at the time of this writing.

Meteor science connects many disciplines: the production, evolution and fate of meteoroids are related to astrochemistry and celestial mechanics; ablation of meteoroids is related to aeronomy and atmospheric science; meteorites and extraterrestrial impacts are related to planetary geology; the mineral water and organic matter embedded in meteoroids are related to astrobiology; the meteoroid impact hazard to space assets is related to aerospace engineering. Of course, as meteoroids are direct products of cometary activity, meteor science is most closely related to comet science and much effort in the past has been in this field. 

In this chapter, we review the development and current state of affairs of the science of meteors in the context of the research on comets. We discuss the process that brings cometary dust to Earth (\S~2) and how we can use this process to study comets (\S~3). The known parent--shower associations are reviewed in \S~4. Interrelations with comet science, astrobiology, aerospace engineering and planetary defense are discussed in \S~5. The chapter is complete with a list of topics into which we hope to see significant progress between now and {\it Comets IV} (\S~6).

\section{\textbf{FROM COMET TO EARTH}}

\subsection{Formation and Evolution of Meteoroid Streams}

Meteoroid ejections are most commonly driven by gas drag from the sublimation of water ice and other volatiles, but can also occur due to break up, impacts, radiation sweeping, and electrostatics gardening, amongst others \citep[cf.][]{Jewitt2015}. The activity mechanism dictates the ejection place, velocity distribution, and size distribution of the meteoroids. For sublimation-driven ejection, the ejection speed grossly follows $\propto a_\mathrm{dust}^{-1/2}$ with $a_\mathrm{dust}$ the diameter of the dust \citep{Whipple1950}. For impulsive ejections such as rotational disruption or impact-caused ejection, observations show that the ejection speed is largely independent of dust size \citep[][Figure 18]{Jewitt2015}. The difference in ejection speed gives the resulting cometary dust trails different looks, allowing activity mechanism to be constrained observationally. This is discussed in detail in the chapter by {\it Agarwal et al.}. Here, we focus on the delivery from comet to Earth and the manifestation of meteor showers at Earth.

Sublimation activity can, in most cases, explain the observed streams (Table~\ref{tbl:shr}), but some streams are associated with parents that have either disrupted or have a history of disruption. It is likely that disruption is the dominant mass loss mechanism. Streams with identified parents only make up 20\% of the confirmed streams. The majority of the confirmed streams do not have known corresponding parents, and most of them -- especially the short-period ones -- are likely the end-product of catastrophic disruptions, given that most sizeable NEOs have been discovered \citep{Jedicke2015}. Taking a typical dispersion timescale of $10^3$~yr for JFC streams (see \S~\ref{sec:comet2earth:demise}) and the number of breakup-driven streams $N=6$ from Table~\ref{tbl:shr}, we derive a breakup frequency of $6\times10^{-3}~\mathrm{yr^{-1}}$ for Earth-crossing JFCs. This is in line with the number derived by \citet{Chen1994} from comet data.

\begin{figure*}[ht!]
\begin{center}
\includegraphics[width=15cm]{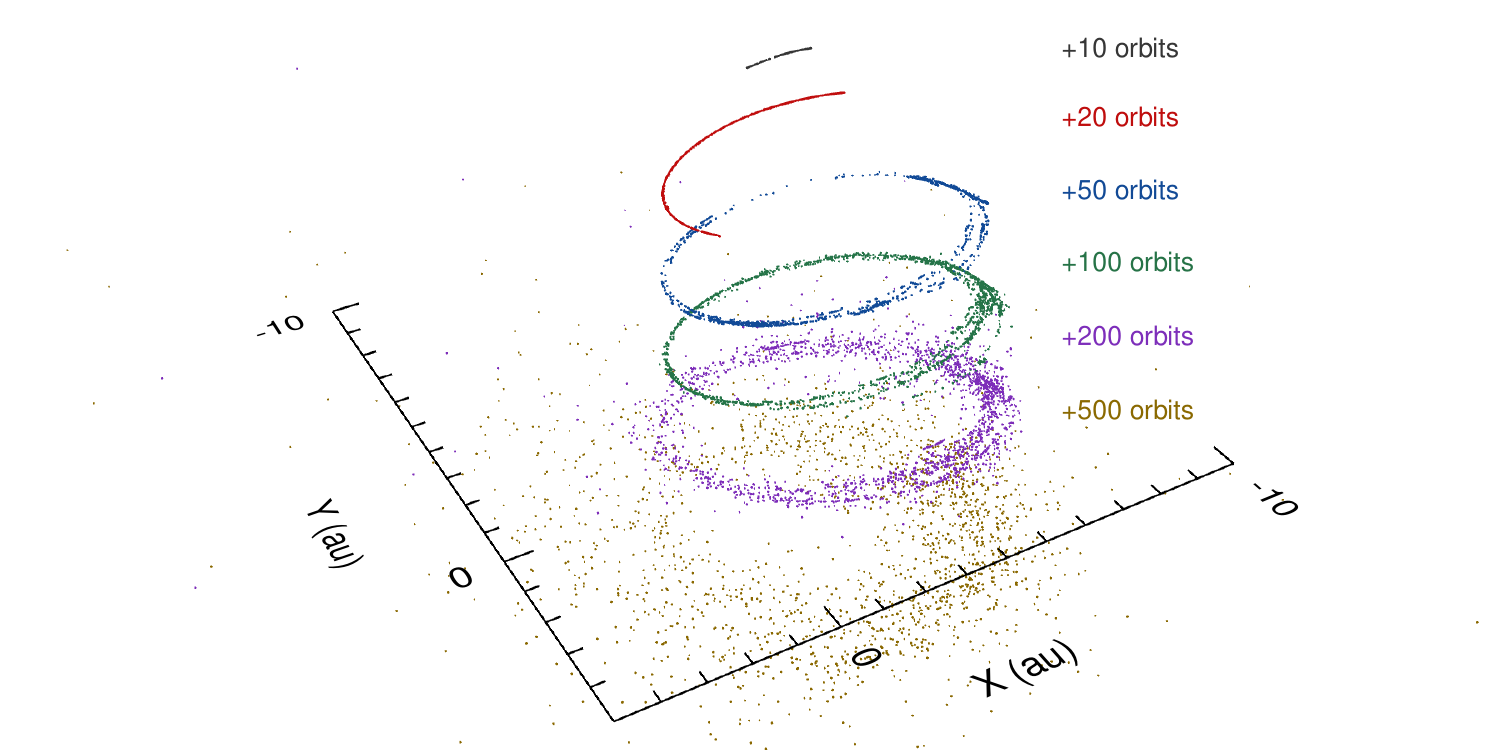}
\caption{Simulation of mm-sized meteoroids ejected by 21P/Giacobini--Zinner during its 1966 perihelion after (from top to bottom) 10, 20, 50, 100, 200, and 500 orbits, showing the evolution of a cometary ejecta from a meteoroid trail to stream, and eventually blending into the interplanetary dust background.
}
\label{fig:evolv}
\end{center}
\end{figure*}

The early manifestation of meteoroid streams is dictated by the fact that meteoroids escape the comet's weak gravity field at a small relative speed with respect to the comet, typically in the order of a few 1--$10~\mathrm{m~s^{-1}}$, which results in a small difference between the orbit of the meteoroid and its parent body. This difference, albeit small, leads to an appreciable difference in the orbital period of the meteoroids. What was ejected as a spherical cloud returns as a trail after even a single orbit. Dynamical processes, such as the gravitational perturbation from major planets, gradually amplified this difference, causing the ejecta to spread along and away from the parent orbit (Figure~\ref{fig:evolv}). This enables meteoroids to arrive at the Earth even when the orbit of the parent is not strictly intercepted by the Earth. In general, comets with minimum orbit intersection distance (MOID) within $\sim0.1$~au can produce observable meteor showers \citep{Jenniskens2021}. It usually takes $\sim10$--20 orbits for meteoroids ejected at a certain epoch to spread to the entire orbit \citep{Ye2016c}.

A meteor shower can be detected annually when the material spreads out to the entire orbit, since the Earth always sweeps some meteoroids every time it passes the intersection with the stream orbit. Alternatively, meteor ``outbursts'' are caused by younger meteoroid trails that reside in a small but growing arc of the orbit. Whether or not that trail meets Earth depends on whether the trail is steered in Earth's path and whether that arc is near Earth. Occasionally, meteoroids can be trapped in resonances if their parents are in the proximity of one of the resonance points, leading to a long-lived meteoroid ``Filament'' or ''trailet''. These trailets can last a few times longer than typical trails, providing a way to probe older materials ejected by the parent. A good example is the Ursid meteor shower which originated from 8P/Tuttle, of which the 7:6 resonance with Jupiter has produced trailets lasting $\sim50$~orbits (\S~\ref{sec:obs:linkages:8p}).

\subsection{Demise of Meteoroid Streams and Contribution to the Zodiacal Cloud}
\label{sec:comet2earth:demise}

All meteoroid streams will eventually disperse into the interplanetary dust complex, known as the ``sporadic (meteor) background''. The coherence of a stream can be accessed using the dissimilarity criterion, $D$, which measures the (dis)similarities between a set of meteoroid orbits based on the difference in their orbital elements \citep[see][\S~9.2.1, for a review]{Williams2019}. A threshold of 0.1--0.2 is typically used to separate streams from non-streams, though difficulty can and has arisen to distinguish weak showers from the noise, especially when the statistics are low. The decoherence timescale varies greatly from stream to stream, and can take as little as $\sim200$~orbits if the stream is heavily-perturbed \citep{Ye2016c}.

Earth's orbital motion gives rise to several very broad apparent radiant regions in the sporadic background: apex sources (meteors arriving from Earth's direction of motion), helion/antihelion (from the direction of the Sun and the opposite direction), and toroidal (particles on prograde orbits with high inclinations). These have been associated with dispersed meteoroids from mostly long-period comets (LPC), Jupiter family comets (JFC) and Halley-type comets (HTC), respectively \citep{Wiegert2009, Nesvorny2010}.

As meteoroid streams lose their coherence, meteoroids also degrade through collision with other meteoroids from the sporadic background. The timescale of this process is uncertain. By combining the then-available meteor, lunar crater and satellite measurements, \citet{Grun1985} found that the collisional lifetime of millimeter-sized meteoroids at 1~au is as short as $10^4$~yr, though recent results from video and radar meteor surveys with much larger statistics suggested $10^5$ to $10^6$~yr, appropriated to meteoroids from sub-millimeter to centimeter sizes \citep{Wiegert2009, Jenniskens2016}. Degradation through other erosion effects, such as thermal fatigue, exposure to solar wind particles and cosmic rays, are possible but are similarly little understood. The timescale that these erosion processes are broadly comparable to the decoherence timescale of JFC streams, but shorter than that of HTC/LPC streams.

Ongoing erosion processes sometimes result in meteor clusters. Only a handful of plausible cases have been reported, mostly from HTCs \citep{Kinoshita1999, Watanabe2003} and one from an unknown LPC \citep{Koten2017}. Dynamical modeling showed that these clusters may be produced a few days before entering the Earth's atmosphere at a separation speed on the order of $0.1~\mathrm{m~s^{-1}}$. The lower tensile strengths of HTCs/LPCs dust likely contribute to the fact that all observed clusters are exclusively from these comets. A recent search in the short-period Geminid shower produces no statistically significant detection \citep{Koten2021}. 

Breakups caused by collision and other erosive processes increase the number of small meteoroids in the sporadic background. About 98\% of small $100~\micron$-class meteoroids are sporadics, but only about 60\% cm-sized meteoroids are not associated with showers \citep[][\S~26.3]{Jenniskens2006}. Collisional processes also steepen the size distribution of the stream, an effect that has been observed in both radar and optical observations \citep{Blaauw2011, Jenniskens2016}. Eventually, meteoroids are reduced into grains that are small enough to be blown out of the solar system.

\section{PROBING COMETS USING METEOR OBSERVATIONS}

\subsection{Observational Techniques}

Meteors and meteoroids can be probed using a wide range of techniques from remote sensing and in-situ exploration in space or on the ground, but most meteors are studied using remote-sensing in optical or radio wavelengths. Optical observations detect meteors using their thermal and ionization radiation, while radar observations detect meteors using the reflectivity of their plasma trail to radio waves. Given the unpredictable nature of meteors, meteor observations are done in the form of ``blind surveys'', and virtually all meteor detections are serendipitous in nature. Most meteor surveys typically aim to collect the trajectory and/or light-curve (or time-amplitude series for radio observations) of a large sample of meteors. A small number of surveys are specially designed for specific purposes such as meteor spectroscopy (which can broadly distinguish chondrite-like vs. non-chondrite-like meteoroids) or high spatial/temporal resolution morphology. Detail reviews of these techniques are given by \citet{Koten2019} and \citet{Kero2019}.

Since ground-based remote sensing techniques detect meteors using the radiation they release/can reflect, there is an observational bias favoring fast energetic meteors, namely the ones originated from HTCs/LPCs. The luminous power is proportional to the kinetic energy of the meteoroid, i.e. $\propto V^2$, hence meteoroids from HTC/LPC sources (with mean arriving speed of $\sim60~$km/s) are around $10\times$ ``brighter'' than JFC ones ($\sim20$~km/s) with the same sizes.

Most optical and radar observations are conducted from the ground, but advances in technology in recent years have enabled observations from airborne or even spaceborne platforms \citep[e.g.][]{Jenniskens1999, Vaubaillon2013, Jenniskens2018}. Compared to ground-based observations, air- and space-borne observations have access to windows at ultraviolet and infrared wavelengths and are not limited by weather, and thus are able to explore new regimes with higher efficiency. Progress is also made in the observation of extraterrestrial meteors in the form of impact flashes and meteoric debris layers, which provide valuable information on the distribution of meteoroids beyond Earth's orbit. Examples include impact flashes on the moon and Jupiter, possible meteor detection by Mars rovers, and the unexpected meteor storm on Mars brought by LPC C/2013 A1 (Siding Spring). Detailed reviews on these topics are given by \citet{Hueso2013}, \citet{Christou2019} and \citet{Madiedo2019}.

Finally, the observation of newly-ejected meteoroids is an important part of any comet mission, which naturally provides knowledge about the birth of a meteoroid stream. This is discussed in detail in the chapter by \textit{Snodgrass et al.} in this volume, which we do not repeat here.

\subsection{Modeling of Meteoroid Streams}
\label{sec:comet2earth:modeling}

Because much of a meteoroid's orbital evolution is determined by the known forces of gravity and radiation pressure, it is possible to predict when meteoroid streams are steered into Earth's path, or when dense sections of the stream are encountered. An important application of meteoroid stream modeling is to predict meteor outbursts and storms, which can guide dedicated observations to study the outburst itself as well as the parent body. \citet[][\S~12 and \S~15]{Jenniskens2006} reviewed the early attempts to predict meteor outbursts. It was not until the Leonid storms of the late 1990s that some meteor outbursts could be reliably predicted. The main difficulty was the lack of computing power to calculate the trajectory of a large number of particles with different orbits. This was solved by a simplified method, developed by \citet{Kondrateva1985} and \citet{McNaught1999}, that significantly reduced the calculation time and successfully predicted the 1999 Leonid storm by dynamical modeling of meteoroid streams.

The subsequent advance of computing technology at the turn of the century allowed a large number (up to many millions) of particles to be simulated. \citet{Vaubaillon2005a, Vaubaillon2005b} incorporated a hydrodynamical model for meteoroid ejection and utilized realistic dust production rate derived from comet observation (e.g. measurement of the $Af\rho$ quantity), and successfully demonstrated their model during the 2002-–2006 Leonids. Those methods have been further developed to study other showers \citep[see][for a review]{Egal2020}. By carefully accounting for the dynamical and physical evolution of the parent comet, the model can generally achieve minute-level accuracy in predicting the timing of the meteor outbursts. It is worth noting that this approach shares many similarities with the technique used by the comet community to simulate cometary tails, except that it works over a much longer timescale (many years) and addresses a different observable (encounter with the Earth, rather than the on-sky distribution of dust). 

\begin{figure*}[ht!]
\begin{center}
\includegraphics[width=15cm]{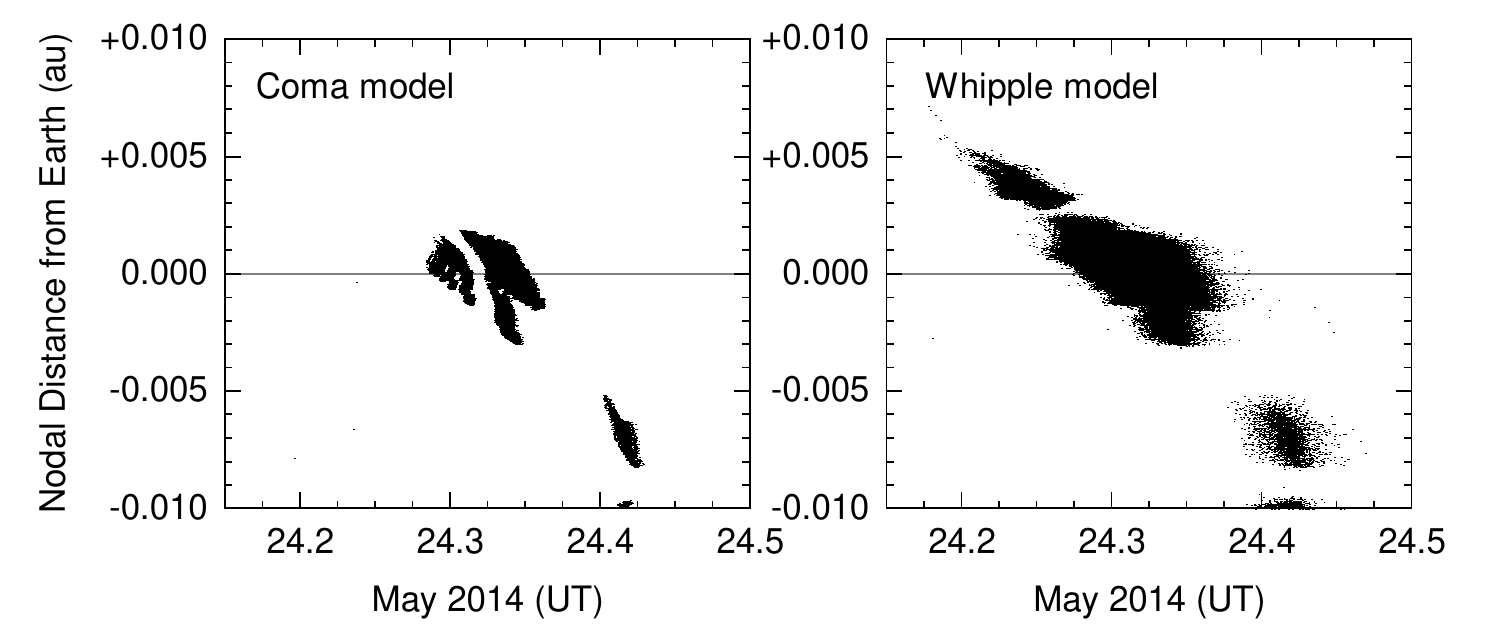}
\caption{Footprints of meteoroid trails from low-activity comet 209P/LINEAR under two different ejection models: the one derived from fitting the observed coma (left panel) vs. the traditional \citet{Whipple1950}'s sublimation model. Meteor observation can provide constraints to the ejection properties of the meteoroids and provide insights into the activity of the parent comet. In this case, a low terminal speed of the meteoroids (left panel) indicates a faster gas-dust decoupling which implies a low volatile content on the nucleus.
}
\label{fig:ej}
\end{center}
\end{figure*}

It is possible to combine the two approaches to achieve a higher degree of realism by replacing the hydrodynamical ejection model with an ejection model derived from comet data. This can be important for certain comets (e.g. low-activity comets) which requires additional tuning to the traditional sublimation theory to accurately describe. In another application, this exercise also allows us to use meteor observations to explore ejection properties of the comet. An example is the observations of low-activity comet 209P/LINEAR and its associated Camelopardalids meteor shower (Figure~\ref{fig:ej}; see also \S~\ref{sec:obs:linkages:209p}).

Predicting the strength of meteor outbursts is more difficult, since it requires knowledge of the historic activity of the parent comet and the ejection conditions. These questions cannot be easily tackled without a full simulation of the entire meteoroid stream, hence early attempts focused on utilizing the meteoroid flux measurements made during historic encounters of the stream in question to project the flux of future encounters \citep[e.g.][]{Jenniskens1998, McNaught1999}. This requires observations of past encounters of the stream, which limits its applicability to a few well-observed streams. 

The initial predictions of the Leonid storms using this approach were only accurate to about an order of magnitude. The ``full'' numerical model, on the other hand, simulates the entire meteoroid stream and therefore can directly translate the dust production of the parent to the fraction of simulated particles arriving at the Earth. In return, the measured flux of the meteor outburst can be used to constrain the historic level of activity of the parent and to tune predictions of future encounters. The accuracy of this quantity is usually within a factor of a few from the measured value, depending on the knowledge of the often-poorly-constrained size distribution of the stream which varies over size range, time and from comet to comet \citep{Fulle2016}.

\subsection{Linking Meteoroid Streams To Their Parents}

The paring of meteoroid streams and parent bodies is started by examining the similarities of their orbits, e.g. by using the $D$-criterion discussed in \S~\ref{sec:comet2earth:demise}. This is not always straightforward, as planetary dynamics constantly perturb the orbits of both the stream and the parent, gradually erasing/altering their dynamical memories. Pairs in stable dynamical ``sweet-spots'' can be traced over a longer timescale (sometimes $\gtrsim 10^3$~orbits for JFCs), while pairs with frequent close encounters with Jupiter can decouple in only a few orbits.

Thanks to the dramatic increase in the orbital data of small bodies and meteoroid streams, many possible parent--stream linkages have been identified in the past two decades \citep[see][Table~7.1 for a summary]{Vaubaillon2019}. However, this also brings the challenge of distinguishing genuine linkages from chance alignments, which is further complicated by the difficulty to obtain precise orbit of meteoroid streams and (sometimes) the parents as well as the fragmentary history of some objects. Uncertainties and confusion can also arise when fragmentation results in multiple parent candidates for a given shower. In those cases, it is not always clear if the fragmentation itself, or later activity from the remaining bodies is responsible for meteor shower activity.

The statistical significance of a linkage can be tested using NEO population models. \citet{Wiegert2004} presented a statistical method that used the \citet{Bottke2002} de-biased NEO population model to calculate the probability of chance alignment of a small body and a meteoroid stream. \citet{Ye2016c} applied this method to all proposed parent--stream linkages and found that only 1/4 of them are statistically significant. 

Another challenge is to confirm cometary parents from the more numerous asteroidal candidates. For instance, the probability of 169P/NEAT being a chance alignment parent to the $\alpha$-Capricornids is either 1 in 3 or 1 in 30 depending on the population being examined -- NEOs or near-Earth comets, while the probability for asteroid 2017 MB$_1$ is 1 in 250 \citep{Ye2018}. A higher significance provides a stronger argument on the ground of statistics, but does not exclusively prove the linkage to be real, hence a proposed linkage needs to be critically examined together with other evidence (e.g. stream modeling).

\section{KNOWN LINKAGES}
\label{sec:obs:linkages}

Here we review the established parent--stream linkages, with a focus on the knowledge of the parent body enabled by meteor data. We note that some of the cases have been recently reviewed by \citet{Kasuga2019}, thus for these cases, we only repeat the fundamental findings.

\subsection{Complexes}

A {\it complex} is a group of dynamically associated parent bodies and meteoroid streams that involve multiple parent bodies and/or multiple streams. As of this writing, four complexes have been established.

\subsubsection{(3200) Phaethon, (155140) 2005 UD and the Phaethon--Geminid Complex}
\label{sec:obs:linkages:3200}

\begin{figure*}[ht!]
\begin{center}
\includegraphics[width=15cm]{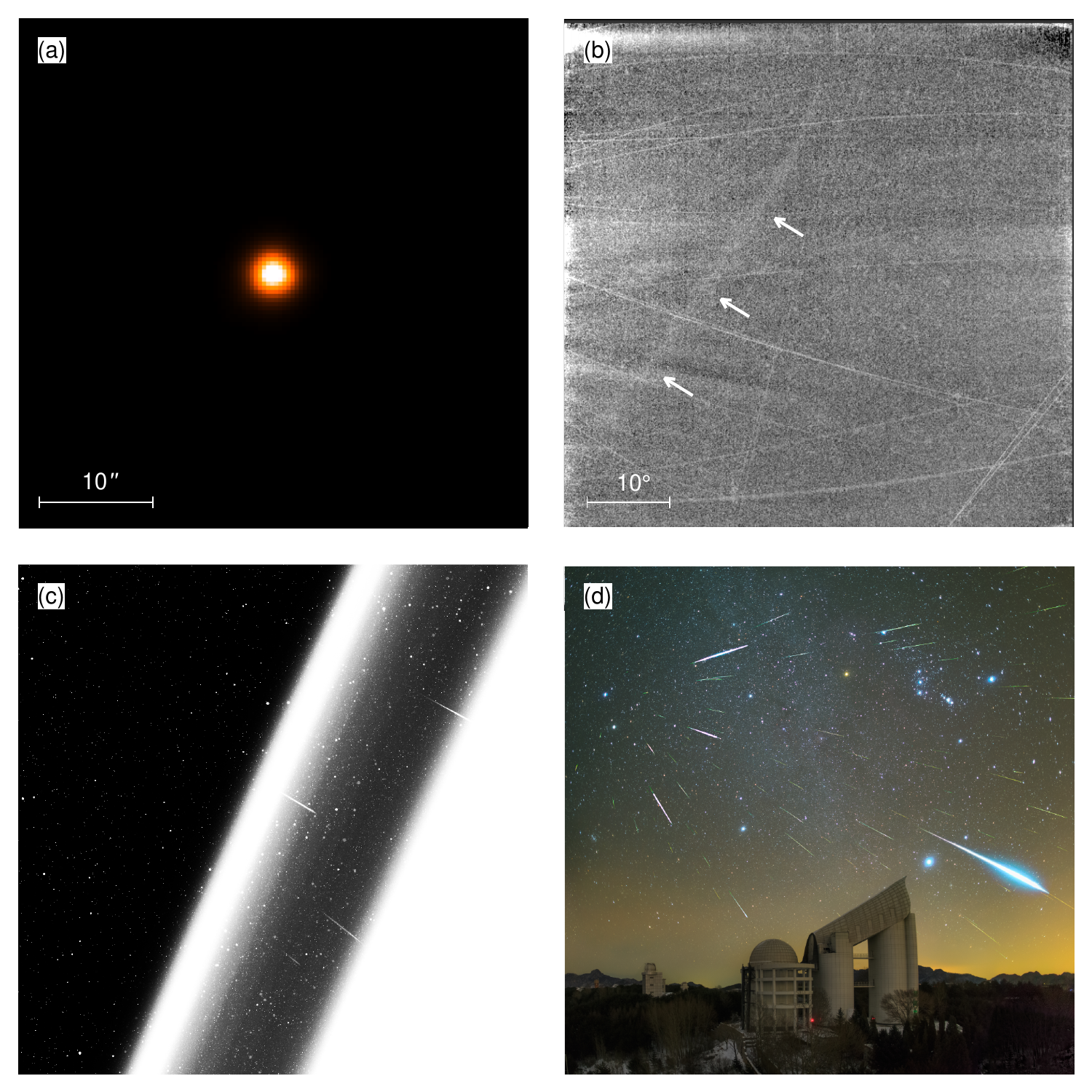}
\caption{Phaethon and various stages of the Geminids: (a) image by the Lowell Discovery Telescope showing an inactive Phaethon \citep{Ye2021b}; (b) the Geminid meteoroid trail in space imaged by {\it Parker Solar Probe} \citep{Battams2022}, marked by arrows; (c) composite image by Yangwang-1 satellite showing atmospheric entry of the Geminids meteoroids; and (d) composite image of the Geminid meteors from the ground, showing the shower radiant (courtesy of Steed Yu).
}
\label{fig:pgc}
\end{center}
\end{figure*}

The Phaethon--Geminid Complex, proposed by \citet{Ohtsuka2006}, includes asteroid--stream pairs (3200) Phaethon -- Geminids (\#4; Figure~\ref{fig:pgc}), (155140) 2005 UD -- Daytime Sextantids (\#221), and possibly asteroid (225416) 1999 YC (\citealp{Jenniskens2006}, \S~22; \citealp{Kasuga2008}; \citealp{Kasuga2009}). These bodies and streams share similar orbital characteristics, being short, relatively highly inclined, and eccentric orbits, with low perihelion that can be explained by an evolutionary sequence by rotating the nodal line (a line defined by the intersection of the asteroid's or stream's orbital plane with the ecliptic plane). They are thought to have originated from a significant fragmentation event \citep{Ohtsuka2006, Kasuga2009}. A connection to (2) Pallas has been proposed \citep{DeLeon2010}, but is debated \citep{MacLennan2021}.

Radar data show that Phaethon is a spinning top rubble pile prone to mass loss by spin-up \citep{Taylor2019}. Observations of Phaethon by the Solar Terrestrial Relations Observatory (STEREO) revealed a $\sim2\times$ brightening a few days around perihelion accompanied by a tail, likely due to thermal fracturing or desiccation of surface material \citep[e.g.][]{Jewitt2010, Li2013}. While Phaethon's linkage to the Geminids is well-established, the mass loss rate for this recurring activity is too small to explain the formation of the Geminids. Recent telescopic observations showed different spectral characteristics between Phaethon and 2005 UD which appears to suggest an independent origin \citep{Kareta2021}, but could also reflect a different history of surface weathering. 

The Geminids are the strongest annual meteor shower with a peak around December 13/14 every year. Its current ZHR is about 180 and has been increasing at a rate of $+20$ per decade \citep[][Figure~4.8]{Koten2019}. The ZHR is expected to reach a plateau of 190 in the next few decades as the core of the stream moves into Earth's orbit \citep[][\S~22.8]{Jones1986, Jenniskens2006}, consistent with the recent upward trend. The formation of the Geminids may have occurred as recently as $\sim1$~kyr ago and coincided with the epoch that Phaethon reached a minimum $q$ in its cyclic orbital variation \citep{Williams1993, MacLennan2021}. Hence, its disruption is likely related to the thermal destruction of near-Sun asteroids \citep{Granvik2016}, but the details remain unclear.

Understanding the dynamics of the Geminid stream has proven to be challenging, mainly due to the difficulty in reconciling the traditional dust ejection theories with the observed timing and duration of the shower. Specifically, various models consistently predict a maximum that is $\sim1$~day later than the observation \citep[cf.][]{Ryabova2016}. A possible explanation, opened up by \citet{Lebedinets1985} and further discussed by \citet{Ryabova2016} and \citet{Kareta2021}, is that the orbit of Phaethon has changed significantly during the formation of the Geminid stream. In other words, meteoroid stream modeling that bases on Phaethon's modern-day orbit may not correctly capture the true evolution of the stream. It then becomes a curious question that how much of our understanding of the dynamical history of the Geminids, Phaethon, and even PGC itself, is still valid \citep{Kareta2021}.

The Geminid meteoroids are exceptional in that they penetrate relatively deep in Earth's atmosphere and rarely show flares \citep[][\S~22.4]{Jenniskens2006}. Spectroscopy of Geminid meteors revealed a large variation in sodium content, with the sodium D-line varying from undetectable to strong \citep{Borovicka2005, Kasuga2005b}. This is likely due to the thermal desorption of some sodium-bearing minerals, of which the mineralogical identity is unclear. \citet{Kasuga2006} showed that the $q$ of the Geminid stream is beyond the distance needed to melt alkaline silicates on meteoroids. Because of the young age of the Geminid stream and the variety of sodium contents, the sodium desorption likely took place when minerals where heated on the surface of Phaethon.

The Daytime Sextantids is a minor daytime shower with ZHR=5 and a peak around September 27 every year. Stream modeling showed that the relative position of the nodal line to 2005 UD demands that the Daytime Sextantids are at least $>10^4$~yr old, which is significantly older than the Geminids \citep{Ohtsuka1997, Jakubik2015}. This suggests that the two showers were formed in separate disruption events and have different parents, instead of both showers originating from the hypothetical breakup that produced Phaethon and 2005 UD \citep[][\S~22.1]{Jenniskens2006}. Running a statistical significance test \citep{Ye2018} on the phase-synchronized orbits of Phaethon and 2005 UD \citep[see][]{Ohtsuka2006}, we find that the likelihood of chance alignment for the Phaethon--2005 UD pair is extremely low -- below 1 in 10,000. However, the issue in the dynamical understanding of the Phaethon--Geminids pair, as discussed above, undermines this conclusion.

\subsubsection{2P/Encke and the Taurid Complex}
\label{sec:obs:linkages:2p}

The Taurid Complex includes comet 2P/Encke, possibly a dozen NEOs, and four established annual showers: the Northern Taurids (\#17), the Southern Taurids (\#2), the Daytime $\beta$-Taurids (\#173), and the Daytime $\zeta$-Perseids (\#172). The daytime showers are active from May to July, while the nighttime showers are active from September to December. 

2P/Encke is known for its bright dust trail in the thermal infrared \citep{Kelley2006}. The comet currently has a significant mass-loss rate of $\sim10^{10}$~kg per orbit. Considering that the total mass of the Taurid Complex ($10^{13}$~kg) and a dynamical age of $\sim10^4$~yr \citep[][\S~25.1]{Jenniskens2006}, this seems to imply that sublimation activity from 2P/Encke alone is sufficient to explain the formation of the Taurid Complex if that activity was as strong as today.
 
The wide dispersion of the streams and decoupling from Jupiter implies a significant age, and early studies have linked a number of NEOs to the complex. By adding up the masses of the meteoroid complex and these NEOs, \citet{Asher1993} suggested that the complex might have been formed by a disrupted 40-km-class comet about $10^4$~yr ago. We since understand that most are O- and S-class asteroids and many of these associations are likely chance alignments, contributed in large part by the low inclination of these objects (\citealp[][\S~25.2]{Jenniskens2006}; \citealp{Egal2021}). 

The northern and southern branches (Northern Taurids and Southern Taurids) are not mirror images in terms of activity and don't represent a full rotation of the nodal line. The four streams may in fact be a sequence of individual showers that were formed more recently than the time it takes to complete a full rotation of the nodal line. A small number of candidate NEOs have since been proposed associated with the complex, all with a semi-major axis very similar to that of 2P/Encke and the low-albedo members in the complex, with the most notable member to be 2004 TG$_{10}$ (\citealp[][\S~25.2]{Jenniskens2006}; \citealp{Jenniskens2016}). Each candidate object appears to link to a period of enhanced Taurid rates. However, some of these objects have short orbital arcs, and hence their connection to the Complex remains to be verified. While the Taurids include large meter-sized objects \citep[and some of these were earlier suspected to produce meteorites, cf.][]{Brown2013}, Encke cometesimals do not appear to be strong enough to survive atmospheric penetration \citep{Devillepoix2021}. 

Spectroscopy of Taurid meteors has shown occasional hydrogen lines (carbonaceous matter), a variable Fe content, and low meteoroid strengths, consistent with a cometary origin \citep{Borovicka2007, Borovicka2019}. 

\subsubsection{96P/Machholz 1, (196256) 2003 EH$_1$ and the Machholz Interplanetary Complex}
\label{sec:obs:linkages:96p}

The Machholz Interplanetary Complex is a gigantic system that consists of 96P/Machholz 1, two near-Sun comet groups, the Marsden group and the Kracht group \citep{Sekanina2005}, asteroid (196256) 2003 EH$_1$, and at least four established annual showers, the Quadrantids (\#10), Daytime Arietids (\#171), and Northern (\#26) and Southern $\delta$-Aquariids (\#5; \citealp[][\S~23]{Jenniskens2006}; \citealp{Babadzhanov2017}). Comet C/1490 Y1 may also be a member of the Complex \citep{Hasegawa1979}. These objects and streams reside in a narrow orbital corridor space and are highly evolved, hence it is challenging to understand the exact relations between these objects and streams \citep{Wiegert2005}.

96P is known to be an outlier among the solar system comet population, with depleted carbon-chain species and a bluer surface \citep{Schleicher2008, Eisner2019}. The Marsden and Kracht comets have only been observed by the space-based Solar and Heliospheric Observatory (SOHO) and are thus poorly understood. Asteroid 2003 EH$_1$ has not been observed to be active \citep{Kasuga2015}, exhibiting a C-like spectrum but with an absorption band around $1~\micron$ due to unknown reasons \citep{Kareta2021}.  

All established showers in this Complex produce significant annual meteor activity. The Quadrantids is characterized by a strong (ZHR=130) but short (only lasts $\sim0.5$~d) maximum, with an annual peak around January 3/4. It is worth noting that observations and dynamical simulations seem to suggest the Quadrantids being a highly variable shower from year to year due to the perturbation from Jupiter \citep[][\S~20.1, Figure~20.17]{Jenniskens2006}. Video and radar data revealed a broad and much weaker component of the Quadrantids underlying the main peak (\citealp{Jenniskens2006}, \S~20.1; \citealp{Brown2010}). the Daytime Arietids is the strongest annual daytime shower, reaching ZHR=60 around mid-June. The two $\delta$-Aquariids are characterized by a relatively broad peak, reaching ZHR=30 around late July. The Southern $\delta$-Aquariids also have an underlying broad component \citep{Jenniskens2016c}. 

Stream modeling showed that 96P or its progenitor is likely the bulk supplier to all of these streams, especially the broad underlying components, composed of meteoroids ejected from 96P about 12--20~kyr ago \citep{Abedin2018}. The Marsden and Kracht comets, likely have originated from a secondary fragmentation 1--2~kyr ago \citep{Abedin2018}, are responsible for some of the Daytime Arietid stream \citep{Sekanina2005}, though their contribution may be shadowed by 96P \citep{Abedin2017}. Asteroid (196256) 2003 EH$_1$ is thought to be the direct parent of the core of the Quadrantid stream. The formation of the Quadrantids core may have occurred as recently as 200--500~yr ago \citep{Jenniskens2004, Wiegert2004b, Abedin2015}.

Despite having a near-Sun history, meteor spectroscopy showed that the Quadrantid meteors are not as sodium-depleted as the Geminids meteors that also have similar history \citep{Borovicka2005, Kasuga2005b}. The underlying cause is unclear \citep{Kasuga2019}, but may be related to the recent formation history during which the perihelion distance was large.

\subsubsection{169P/NEAT and the $\alpha$-Capricornid Complex}
\label{sec:obs:linkages:169p}

The $\alpha$-Capricornid Complex consists of JFCs 169P/NEAT, P/2003 T12 (SOHO), asteroid 2017 MB$_1$, and the $\alpha$-Capricornid shower (\#1). The orbits of 169P and P/2003 T12 are almost identical, and dynamical investigation shows that the spatial distance and relative speed of the pair reached a minimum 2900~yr ago, suggestive of a breakup origin \citep{Sosa2015}. Following the approach of \citet{Ye2018}, we estimate that the likelihood of chance alignment is 1 in 10,000, a remarkably small number. On the other hand, as will be discussed below, there is an appreciable difference between the orbits of the 169P--P/2003 T12 pair and 2017 MB$_1$, hence it is not clear if all these three bodies are indeed dynamically related. Additional linkages to working-list showers $\xi^2$-Capricornids (\#623) and $\epsilon$-Aquariids (\#692) have been proposed \citep{Jenniskens2016c} but are in need of further investigation.

The $\alpha$-Capricornids reaches a modest ZHR=5 in late July and is characterized by an abundance in bright meteors. Numerical modeling suggests that the currently observed $\alpha$-Capricornids were ejected 4500--5000~yr ago by a major disruption event \citep{Jenniskens2010, Kasuga2010}. More recent ejecta have not evolved to the Earth's orbit to be observed, hence the comet must have been much more active in the past than the current mass-loss rate of $\sim10^{-2}$~kg/s reported by \citet{Kasuga2010}.

Recently, NEO 2017 MB$_1$ was found to have an orbit closer to the mean $\alpha$-Capricornid orbit \citep{Wiegert2017}. The likelihood of chance alignment of the $\alpha$-Capricornids--2017 MB$_1$ pair is 1 in 250, a lot lower than that of 169P and P/2003 T12 (1 in 3 and 1 in 5, respectively). It is possible that 2017 MB$_1$ is one of the macroscopic fragments produced during 169P's major disruption $\sim5$~kyr ago, similar to the case of 96P and 2003 EH$_1$ discussed above. Contrary to 169P and P/2003 T12 which have been observed to be active, 2017 MB$_1$ appears to be inactive. More study is needed to understand the relation between these bodies.

Spectral observations of the $\alpha$-Capricornid meteors revealed their composition being Mg-rich and Fe-poor compared to typical chondrites \citep{Madiedo2014b}. The same authors also showed a relatively high end height of these meteors, which indicates a low tensile strength compatible with a cometary origin.

\subsection{Typical JFCs}

\subsubsection{7P/Pons--Winnecke and the June Bootids}
\label{sec:obs:linkages:7p}

The June Bootids (\#170) exhibited significant activity in 1916, 1998 and 2004, with peak ZHR reaching 100, but activity is nearly nonexistent in other years \citep{Arlt2000, Jenniskens2004e}. Meteor activity in 1998 and 2004 occurred when its parent body, 7P/Pons--Winnecke, had evolved away from the Earth ($q=1.26$~au in 1998). The erratic behavior of June Bootids is largely due to the rapid orbital evolution of 7P. The comet's frequent close encounter with Jupiter has moved its orbit from $q=0.77$~au from 1809 to $q=1.24$~au currently, passing the Earth's orbit by $0.03$~au around 1916. Stream modeling shows that the June Bootids outbursts in 1916, 1998 and 2004 precipitated from meteoroids ejected in 1819--1869 locked in the 2:1 resonance with Jupiter \citep{Tanigawa2002, Jenniskens2004e}. No encounter with the June Bootid trails is expected in the next few decades, but 7P will return to the near-Earth space following a close encounter to Jupiter in 2037, hence a resurrection of the June Bootids in the future is possible.

A cometary dust trail was detected in the orbit of 7P by the Infrared Astronomical Satellite \citep[IRAS;][]{Sykes1992}. The mass loss rate of 7P can be estimated using the activity of June Bootids. The 1998 June Bootids outburst was dominated by the dust ejected by 7P in 1825 and reached a peak flux of $0.12\pm0.02~\mathrm{km^{-2}~hr^{-1}}$ appropriated to millimeter-sized meteoroids \citep{Brown1998b}, translating into a mass loss of $6\times10^8$~kg per orbit, in agreement with the value determined from photometric variation of 7P \citep{Kresak1987}. Hence, it only takes a few orbital revolutions to accumulate a sufficient amount of dust to be detected by IRAS.

\subsubsection{8P/Tuttle and the Ursids}
\label{sec:obs:linkages:8p}

8P/Tuttle is technically a JFC even though its semi-major axis is considerably larger than most of the JFCs discussed in this chapter ($a=5.7$~au vs. $a\sim2-3.5$~au). The associated meteoroid stream, the Ursids (\#15), is an annual stream known for its ``far-type'' outbursts -- brief heightened meteor activity when 8P is near aphelion. The shower reaches a maximum around December 22 and typically has ZHR=10, but can exceed ZHR=100 during outbursts. Stream modeling shows that this is due to the mean-motion resonances (see \S~\ref{sec:comet2earth:modeling}) which caused the meteoroid clumps to slowly drift away from the position of the comet \citep{Jenniskens2002}. Currently observed outburst meteoroids were released by 8P in 9th--14th century \citep[][Table~5a]{Jenniskens2006}. By comparing the stream model and modern observation of the comet, it can also be inferred that the activity of 8P has not changed appreciably over the past millennium (100~orbits). The outburst Ursids exhibited higher begin/end heights compared to the background Ursids \citep{MorenoIbanez2017}, indicative of higher fragility typically seen in cometary meteoroids. 8P currently has MOID=0.1~au to the Earth, hence more recently released meteoroids will not reach the Earth in the near future as long as the stream is not highly perturbed.

\subsubsection{15P/Finlay and the Arids}
\label{sec:obs:linkages:15p}

The Arids (\#1130) is an emerging meteor shower that originated from 15P/Finlay. 15P has an Earth MOID of only 0.0097~au, but the associated meteor activity had not been observed until 2021. \citet{Beech1999} suggested that the lack of a meteor shower was due to the facts that the planetary perturbation that has driven most meteoroids away from the Earth's orbit, and that the comet has not been very active. Recent stream modeling found that ejecta from 1995 to 2014 apparitions of the comet would cross Earth's path in 2021 \citep{Ye2015} which was confirmed by video networks and radar \citep{Jenniskens2021b}. Remarkably, the detected meteor activity was partially contributed by the two comet outbursts detected during its 2014 return \citep{Ye2015, Ishiguro2016}. This is the first time that heightened meteor activity can be traced back to telescopically-detected cometary outbursts. Calculation by Mikhail Maslov (unpublished data) suggests more meteor activity is possible in the coming decades, especially in 2047 when the Earth directly encounters the 2008 trail.

\subsubsection{21P/Giacobini--Zinner and the October Draconids}
\label{sec:obs:linkages:21p}

Although the activity is nearly nonexistent in most years, the October Draconids (\#9; sometimes being referred to as simply the ``Draconids'') has produced some of the strongest meteor activities ever observed in modern history. Meteor storms have been detected in the optical (1933, 1946, 1998) and by radar (1999, 2012), primarily due to the mass-dependent delivery of meteoroids under the dynamical effects \citep{Egal2019}.

Early efforts to model the October Draconid stream have encountered difficulties to reproduce the timing of the 2005 and 2012 outbursts \citep{CampbellBrown2006, Ye2014b}, likely due to the orbital uncertainty introduced by 21P's multiple close encounters with Jupiter and/or variations in nongravitational acceleration. \citet{Egal2019} showed that this discrepancy can be largely overcome by using an orbital solution appropriated to each apparition rather than the solutions averaged over several apparitions. This highlights the challenge of modeling meteoroid streams from comets with highly variable activity and/or have frequent close encounters with major planets. However, the same effect can also potentially provide a powerful tool to probe the dynamics of the comet in the past, especially over times when the comet was not observed telescopically.

The October Draconid meteors are known for their unusually high ablation altitude and abundant fragmentation which can be explained by the extreme fragility of the meteoroids \citep{Jacchia1950}. The begin and end heights of these meteors are typically 10~km higher than other meteors with the same entry speeds \citep[][Figure~10]{Ye2016}. The fragile nature of the meteoroids may be related to the unusual bluing of the dust coma seen in polarimetry \citep{Kiselev2000}. Stereoscopic and spectroscopic observations have revealed physical and chemical heterogeneity of these meteors. The sizes of meteoroid constituents appear to vary appreciably over a factor of 5, but the bulk meteoroids are all very porous with porosities around 90\% \citep{Borovicka2007}. The chemical abundances of the meteoroids are nearly chondritic. Temporally resolved spectroscopy of a bright October Draconid meteor (with an initial mass of $\sim5$~kg) revealed an early release of sodium in the meteoroid ablation which could indicate the presence of hydrated minerals in the meteoroid \citep{Madiedo2013}.

\subsubsection{26P/Grigg--Skjellerup and the $\pi$-Puppids}
\label{sec:obs:linkages:26p}

Similar to June Bootids, the $\pi$-Puppids (\#137) is a shower with minimal activity in normal years but have exhibited occasional outbursts, notably in 1972, 1977, 1982, and 2003 \citep[][\S~19.1]{Jenniskens2006}. Its parent comet 26P/Grigg--Skjellerup experiences frequent close encounters with Jupiter, which changed its $q$ from $0.73$~au in 1808 to the present $1.12$~au. As a result, the $\pi$-Puppids stream has been split into many filaments. There are some discrepancies between the stream model and the observations that are difficult to reconcile \citep{Vaubaillon2005c}, possibly due to the limited observation or the poor knowledge of the filaments. 26P will stay beyond Earth's orbit until a close encounter to Jupiter in 2118. Before that time, Earth might still be able to encounter filaments produced by past activity of the comet \citep[][Table~6e]{Jenniskens2006}.

\subsection{Low Activity JFCs}
\label{sec:obs:linkages:low-activity}

\subsubsection{209P/LINEAR and the Camelopardalids}
\label{sec:obs:linkages:209p}

Despite its relatively large size (diameter $\sim3$~km) and stable orbit in the near-Earth space, 209P/LINEAR was not discovered until 2004. Subsequent telescopic observation of the comet revealed a dust-rich comet with an extremely low activity level \citep{Ye2014, Schleicher2016}. Modeling of the dust coma showed a dust terminal speed that is $10\times$ lower compared to the Whipple model \citep{Ye2016}, likely due to fast dust-gas decoupling caused by low ice content in the dust grains. Dynamical investigation shows that the comet is in a semi-stable orbit, with no close encounter to Jupiter over the past $\sim10^4$~yr \citep{Fernandez2015, Ye2016}.

Meteor activity from 209P, the Camelopardalids (\#451), was first detected in 2014 when the comet made a close approach to the Earth, and a handful of trails generated between 1798 and 1979 were directly crossed by the Earth \citep{Ye2014}. Characterization of the optical-sized meteors showed low tensile strengths compatible with a cometary origin, though spectroscopy also revealed a low Fe content indicative of non-chondrite material \citep{Madiedo2014}. Interestingly, the radar-sized meteoroids are apparently of higher tensile strength \citep{Ye2016}, implying a difference meteoroid structure compared to comets such as 21P (which meteoroids are consistently fragile regardless of size). A previously unnoticed minor outburst in 2011 was discovered by \citet{Ye2016}. Stream modeling showed that the 2011 outburst was produced by trails generated in 1763 and 1768. The observed meteoroid flux is $100\times$ larger than the number calculated using the current dust production rate of 209P, implying an elevated activity of the comet back in those years. Overall, the Camelopardalids is nearly undetectable in most years, showing that 209P has supplied little dust over the past a few kyr, the dispersion timescale of the meteoroid stream. This evidence supports the idea that the comet is about to become an inert object. The Camelopardalid stream, like its parent, is highly stable, and occasional minor outbursts are expected throughout this century.

\subsubsection{300P/Catalina and the June $\epsilon$-Ophiuchids}
\label{sec:obs:linkages:300p}

Radar observation shows that 300P ejects large, cm-class meteoroids \citep{Harmon2006}. A linkage to the June $\epsilon$-Ophiuchids (\#459) has been proposed \citep{Rudawska2014}. This very weak shower was initially only detected through a $D$-criterion clustering search in video data, but a more significant (though still minor) outburst was detected in 2019 \citep{Matlovic2020}. Stereoscopic and spectroscopic observations revealed a chondritic, porous structure of the meteoroids consistent with a cometary origin \citep{Matlovic2020}. The meteors exhibited a relative depletion in sodium, a puzzling feature given that 300P has not been in a low-$q$ orbit in the past several kyr.

\subsection{JFCs --- Fragmentation and Breakup}

\subsubsection{3D/Biela and the Andromedids}
\label{sec:obs:linkages:3d}

The disruption of 3D/Biela is the first well-studied comet fragmentation event in the telescopic era. The comet was observed to start disintegration during its 1846 apparition and was not seen again after 1852. Instead, meteor storms from the associated Andromedid meteor shower (\#18) were seen in 1872 and 1885. Numerical modeling confirmed that these exceptional activities were produced by continuous disintegration of 3D \citep{Jenniskens2007}. However, the same model also suggested that the observed Andromedid stream only accounts for a few percent of the estimated mass of 3D/Biela, prompting the question about the fate of the rest of the mass. One possibility is that one or more macroscopic fragments have survived the disruption and are now inert, but the lack of detection of such objects seems to disagree. Other possibilities include overestimated nucleus size or underestimated stream mass. The current estimate of the nucleus diameter of $\sim3$~km is based on the nongravitational acceleration and brightness derived from the astrometric and photometric data assuming a typical nucleus behavior \citep{Babadzhanov1991}. However, we now know that some small comets can exhibit large nongravitational acceleration and/or high brightness, such as 252P/LINEAR described above. Hence, it is likely that the pre-breakup nucleus of 3D is smaller, perhaps around 0.5~km in diameter if all of the mass has been converted to the meteoroid stream. The Andromedid stream itself remains detectable \citep{Wiegert2013}, albeit mostly at low activity level as the bulk of the stream has been evolved away from Earth's orbit. Planetary perturbation may bring the stream back to the Earth's orbit after 2120 \citep{Babadzhanov1991}.

\subsubsection{73P/Schwassmann--Wachmann 3 and the $\tau$-Herculids}

73P/Schwassmann--Wachmann 3 is known for its spectacular fragmentation in 1995 and 2006. The associated $\tau$-Herculid meteor shower (\#61) was first detected in 1930, shortly after the discovery of the comet \citep{Nakamura1930}. The shower exhibited elevated activity in its 1930 return but has been quiet in other years. Elevated meteor activity is expected in 2022, 2049 and 2065, due to direct encounters with meteoroids ejected in the 20th century, including the interesting 1995 apparition. Models are currently divided on whether the 1995 breakup of 73P can produce meteor storms in 2022 and 2049 \citep{Wiegert2005, Horii2008}, mainly due to the unknown details of the fragmentation. (Non-)detection of a storm will help refine the details of the 1995 breakup.

Detection of a dust trail from 73P has been reported in Cosmic Background Explorer (COBE) data \citep{Arendt2014} as well as dedicated observation by Spitzer \citep{Vaubaillon2010}. Taking ZHR=2 and a shower duration of 0.5~day of the associated $\tau$-Herculid shower, we obtain a stream mass of $5\times10^8$~kg. This does not include the mass deposited by the 1995 and 2006 fragmentation of 73P, which may have produced a similar amount of material \citep{Ishiguro2009}.

\subsubsection{289P/Blanpain and the Phoenicids}
\label{sec:obs:linkages:289p}

289P/Blanpain was originally discovered in 1819 and was subsequently lost (designated as D/1819 W1 for that apparition). More than a hundred years later, following a significant outburst of the then-little-known Phoenicid meteor shower (\#254) in 1956, \citet{Ridley1963} suggested that the meteor shower was related to the lost comet D/1819 W1 (Blanpain). Finally, asteroid 2003 WY$_{25}$ was discovered in November 2003 and was subsequently linked back to D/1819 W1 \citep{Jenniskens2005}, receiving the permanent designation as 289P. It remains unclear whether 2003 WY$_{25}$ is D/1819 W1 itself or a fragment of it. The mass of 2003 WY$_{25}$ is only $\sim1/10$ of that of the Phoenicid stream \citep{Jenniskens2005}, hence it is possible that either more macroscopic objects remain to be discovered, or one or more catastrophic mass-loss events has occurred in the recent history. The former scenario appears to be unlikely given the high completion rate of NEO surveys. 

2003 WY$_{25}$ is still weakly active \citep{Jewitt2006} and has produced a significant ($\Delta m = -9$~mag) outburst in 2013 while near aphelion \citep{Ye2019b}, one of the largest comet outbursts ever observed. The outburst is remarkable considering 289P's small size \citep[a nucleus radius of 100--160~m][]{Jewitt2006}. The outburst accounted for $\sim1\%$ of the remaining mass of 289P and is likely driven by runaway sublimation of supervolatiles triggered by rotational spin-up. A significant nongravitational motion was detected in 289P's subsequent return in 2020 which may be caused by the 2013 mega-outburst. Minor enhancements of the Phoenicid shower have been observed in 2014 and 2019 \citep{Fujiwara2017, Roggemans2020}. A small amount of the ejecta from 2013 outburst will reach the Earth in 2036 and 2041.

\subsection{Comets Without Showers}
\label{sec:obs:linkages:parents}

\paragraph{46P/Wirtanen.} This comet was recently brought into the near-Earth space due to close encounters with Jupiter in 1972 and 1984. Meteor activity has been predicted but not detected \citep{Maslov2017}. Additionally, no dust trail has been reported despite repeat infrared observations by IRAS, the Infrared Space Observatory (ISO) and Spitzer. A faint optical trail was recently detected in the images taken by the Transiting Exoplanet Survey Satellite \citep[TESS;][]{Farnham2019}, providing a lower mass limit of $2\times10^5$~kg. Meteor activity has been predicted but no detection has been reported. Taking a typical detection limit of $\sim10^{-4}~\mathrm{km^{-2}~hr^{-1}}$ of mm-class meteoroids, achievable by routinely-operated meteor surveys nowadays, and a shower duration of 0.5~day, we obtain an upper stream mass limit of $5\times10^7$~kg. On the other hand, high-cadence observations have revealed at least 6 minor outbursts of 46P during its 2018 apparition, each producing $10^4$--$10^6$~kg of material \citep{Kelley2021}. If 46P has been active at this level during its past perihelion passages since arriving at its current orbit in 1984, it should have deposited $10^5$--$10^8$~kg of material along its orbit. The absence of a meteor shower appears to argue for a smaller mass loss rate in this range.

\paragraph{103P/Hartley 2.} Similar to 46P, this comet arrived in the near-Earth space after a close encounter with Jupiter in 1971. It has a similar MOID as 46P (both are 0.07~au) but no meteor activity has been detected. Trail mass constrained from the Wide-field Infrared Survey Explorer (WISE) data \citep{Bauer2011} places it at a similar level as the trail of 2P. A cursory stream simulation suggested a lack of direct encounters to the dust trail until the 2060s \citep[][\S~19.8]{Jenniskens2006}. 103P is known to produce large ejecta of up to 0.2--2~m in sizes \citep{Kelley2013}.

\paragraph{107P/Wilson--Harrington.} This object is known for its singular episodic activity during its discovery in 1949. A linkage to the unconfirmed September $\gamma$-Sagittariids has been proposed by \citet[][\S~9.3]{Jenniskens2006}, a shower that has only been detected once. Significant nondetection of an associated trail has been reported in optical \citep{Ishiguro2009} and infrared \citep{Reach2007}. Following the discussion for 46P, we place an upper limit of $5\times10^7$~kg of any meteoroid stream. 107P, however, is dynamically more stable than 46P as it is quasi-decoupled from Jupiter, hence material from low-level activity (or occasional episodic ejection) has sufficient time to accumulate along its orbit. The lack of meteor activity may indicate a rather infrequent, or even a one-off, ejection.

\paragraph{249P/LINEAR.} The comet is among the most stable cometary objects in the NEO population that may originated in the asteroid belt \citep{Fernandez2015, Fernandez2017}. Meteor activity is possible in mid April (from $\alpha=207^\circ$, $\delta=-19^\circ$ at geocentric speed $v_\mathrm{g}=26.1$~km/s) and early November (from $\alpha=217^\circ$, $\delta=-6^\circ$ at $v_\mathrm{g}=26.2$~km/s), but has not been detected. Further study is needed to understand the visibility of the 249P meteoroid stream and its implication of the past activity of the comet.

\paragraph{252P/LINEAR.} Though originally considered as a low-activity comet, 252P's close approach to the Earth in 2016 (at 0.036~au) coincides with a revival in activity, though primarily in form of gas emission \citep{Li2017}. Meteoroid delivery to the Earth is possible, but no meteor has been detected \citep{Ye2016c}. The comet has a small nucleus with a radius of $300\pm30$~m, but has exhibited strong nongravitational motion during its 2016 return, possibly due to a strong jet \citep{Ye2017}. The nongravitational motion may have evolved the comet away from its likely tiny meteoroid stream, but this possibility has not been thoroughly investigated. Future meteor observation may also help understand certain polarimetric features detected in comet data that have been interpreted as signatures of large, compact grains ejected from a desiccated surface \citep{Kwon2019}.

\paragraph{P/2021 HS (PANSTARRS).} This comet has one of the smallest steady-state active area ever measured, equivalent to a single circular pit with a radius of 15~m \citep{Ye2022}. With a MOID of 0.04~au, meteor activity is possible but has not been detected, placing an upper limit of $5\times10^7$~kg of the stream \citep{Ye2022}. The same authors also reported an independently-derived upper limit of $(3-5)\times10^7$~kg of the stream based on the non-detection of dust trail in the TESS images. Dynamical investigation showed that the comet had a close encounter with Jupiter in 1670, by only 0.05~au, and hence is dynamically unstable as most JFCs. Its extreme depletion of volatile may indicate a highly thermally processed nucleus.

\paragraph{(3552) Don Quixote.} With a diameter of 18~km, Don Quixote is the largest NEA on a comet-like orbit, and has long been suspected to be a dormant comet. Recent observations with Spitzer and ground-based optical telescopes have revealed recurring activity \citep{Mommert2014,Mommert2020}. The cause of the activity, however, is inconclusive. Don Quixote's current MOID to Earth is 0.34~au, much larger than typical MOIDs of meteor-producing parents, but some meteoroids may still evolve into Earth-approaching orbits within a few kyrs \citep{Rudawska2015}.

\subsection{HTCs}

With the exception of 1P/Halley, most known Halley-type comets (HTC) have only been observed once or twice since the dawn of modern astronomy and are poorly characterized. Interestingly, three out of the four streams with well-established parents are associated with large and very active comets that have been sighted throughout a significant part of human history. Whether this is due to observational bias or is a reflection of the physical and/or dynamical evolution of the HTC population in general remains to be investigated. Using meteor spectroscopy and by combining the result with in-situ exploration, it has been recognized that HTC meteoroids are more consistent with the anhydrous, carbon proto-CI material that represents typical cometary dust \citep{Rietmeijer2000, Borovicka2004}.

\subsubsection{1P/Halley, the $\eta$-Aquariids and the Orionids}
\label{sec:obs:linkages:1p}

As an active and long-lived comet that periodically visits the inner solar system, 1P/Halley has produced a complex of meteor showers. The most significant activity are the Orionids (\#8) in late October at the ascending node and the $\eta$-Aquariids (\#31) in early May at the descending node. Both showers are among the strongest annual showers: $\eta$-Aquariids has a ZHR of 60 and the Orionids is about 20. These two showers have been identified since ancient times \citep{Zhuang1977}.

The Earth crosses the two branches at different nodal distances. Currently, the ascending node represented by the Orionids lies farther away from Earth's orbit than the descending node represented by the $\eta$-Aquariids (0.17~au vs 0.06~au), therefore the Orionids may represent older and more evolved materials from 1P. Sampling of material at different distances to the node also provides means to explore the structure of a meteoroid stream. The attempt to model the activity profile of both showers led to the realization of \citet{McIntosh1983}'s shell model, an incomplete rotation of the nodal line due to liberating about a mean-motion resonance, which correctly describes the structure of meteoroid stream. Compared to the $\eta$~Aquariids, the older Orionids seem to have lower tensile strength and have a more symmetric activity profile \citep{Egal2020b}. Both showers show a periodicity of 11--12~yr due to the perturbation of Jupiter, and elevated activity is expected for $\eta$-Aquariids in 2023/24 and 2045/46 \citep{Egal2020c}.

An interesting feature of these streams is the abundance of small, sub-millimeter-class meteoroids which differs from many other meteoroid streams. High power radar observations have confirmed the unusual richness in $<100~\micron$-class meteoroids of the Halleyid stream compared to other streams, including HTC streams \citep{Kero2011, Schult2018}. This appears to be the intrinsic character of 1P, as a similar abundance of small meteoroids was also detected by the Giotto spacecraft during its visit to 1P in 1986 \citep{McDonnell1986}. Spectra of the Orionid meteors closely resemble those of the Perseids, meteoroids from HTC 109P/Swift--Tuttle \citep{Halliday1987}. High-resolution spectrum ($R\sim10000$) of a Orionid meteor has revealed many metal species such as Ti, Cr, Zr, Pd and W \citep{Passas2016}.

\subsubsection{55P/Tempel--Tuttle and the Leonids}
\label{sec:obs:linkages:55p}

The Leonids (\#13) has produced most of the observed meteor storms in modern history, with the most recent ones in 1999--2002. Leonid storms were recorded as far back as 902 AD \citep{Hasegawa1996}, which helps establish the orbital history of the parent comet, 55P/Tempel--Tuttle, beyond its first recorded detection in 1366. The Leonid storm in 1966 may have reached a ZHR of 15,000 \citep{Jenniskens1995}, the strongest meteor outburst ever measured. In normal years, the Leonids are a moderately active shower, reaching ZHR=15 around November 17 each year. The next cluster of strong Leonid activity may arrive around 2031 when 55P returns to perihelion.

Studies of the Leonids have brought two important landmarks in meteor science: understanding of the nature of meteor showers and their relation to comets (\S~\ref{sec:intro}) and successful modeling and prediction of meteor storms (\S~\ref{sec:comet2earth:modeling}). The Leonid Multi-Instrument Aircraft Campaign organized during the 1998--2002 Leonid returns deployed a unique array of new observing techniques and has greatly enhanced our understanding of meteor showers, with mid-infrared and sub-millimeter meteor observations being conducted for the first time \citep[see the reviews by][]{Jenniskens1999, Jenniskens2000c}.

A rather unusual detection of a dust trail from 55P was made by \citet{Nakamura2000} who detected the optical glow of the 1899 trail using a wide-angle lens, when the trail passed very close to the Earth in 1998. Different from the typical trails in telescopic data, this detection manifested in form of a large (a few degrees in radius), diffuse, and largely circular structure in the night sky, which is effectively the cross-section of a meteoroid trail. They derived a radius of $0.01$~au and a number density of $1.2\times10^{-10}~\mathrm{m^{-3}}$. Assuming a toroidal-like structure of the stream with the uniform spatial distribution of particles of $10~\micron$ in size (compatible with the optical detection of the glow) and a bulk density of $400~\mathrm{kg~m^{-3}}$ \citep{Babadzhanov2009}, we derive a total mass of $3\times10^{10}$~kg for the 1899 trail. Given a typical dispersion timescale of 100~orbits (\S~\ref{sec:comet2earth:demise}), this agrees well with the total stream mass of $5\times10^{12}$~kg derived by \citet{Jenniskens2000b}.

Stereoscopic and spectroscopic observations of Leonid meteors revealed a different behavior compared to the other two well-studied HTC showers, the Orionids and the Perseids. The Leonid meteoroids are more fragile, less abundant in carbonaceous matter, and are slightly more abundant in Na and Mg \citep{Borovicka1999, Kasuga2005}. The fragile structure makes the Leonids more similar to the October Draconids (a JFC stream) than to the Perseids. Spectroscopy of the rare meteoric afterglow, the long-lasting visible debris train left by bright fireballs, showed many metal lines such as Fe, Mg, Na, Ca, Cr, Mn, K, and possibly Al in the debris \citep{Borovicka2000}.

\subsubsection{109P/Swift--Tuttle and the Perseids}
\label{sec:obs:linkages:109p}

The Perseids (\#7) is perhaps the most observed meteor shower in history, with sightings dating back to the first millennium. It is the strongest annual meteor shower with a comet parent, reaching ZHR=80 around August 12 of each year, but outbursts up to ZHR=500 have been observed during the return of its parent comet 109P/Swift--Tuttle in 1989–-1995 \citep{Jenniskens1995}.

With an effective diameter of 26~km \citep{Lamy2004}, 109P is the largest known object in the near-Earth comet population. The Perseid meteoroid stream is among the most massive stream crossing the Earth's orbit, with a total mass of $\sim4\times10^{13}$~kg \citep{Jenniskens1994}. A dynamical model constructed by \citet{Brown1998} showed a stream age of $\sim10^5$~yr with a slightly younger core of about $(2.5\pm1.0)\times10^4$~yr old. The elevated activity near the perihelion year is primarily due to the young meteoroids ejected in the last few orbits. Stream modeling shows the age of the outburst material is clearly correlated to the distance to the comet: the outbursts within 2--3 years from perihelion were primarily caused by the youngest meteoroids ejected in the last 1--2 orbits, outbursts about 3--4 years were caused by meteoroids ejected about 3--4 orbits ago, and the ones further away from the perihelion were the older materials. Recent studies have also identified a 12-year cycle of the Perseid activity due to the perturbation of Jupiter, which produces weaker outbursts even further away from the perihelion year \citep[][\S~17.9]{Jenniskens2006}. The last outburst in this cycle occurred in 2016 and reached ZHR=200 \citep{Rendtel2017b}. The next outburst is expected in 2028.

Spectroscopic observations showed that the composition of Perseids are more similar to the dust sampled at 1P than to chondrites \citep{Borovicka2004, Spurny2014}. The meteoroids are depleted in certain metal elements like Fe, Cr and Mn but are enhanced in Si and Na. The abundance in Na may reflect the fact that Na being the interstitial fine-grain material that joins large mineral grains \citep{TrigoRodriguez2007}.

\subsubsection{C/1917 F1 (Mellish) and the December Monocerotids}
\label{sec:obs:linkages:1917f1}

The December Monocerotids (\#19) is a modest annual stream active in mid-December characterized by its relatively small perihelion of $q=0.19$~au. A connection to near-Sun comet C/1917 F1 (Mellish) was first proposed by \citet{Whipple1954} and supported by recent meteor data \citep{Lindblad1990, Veres2011}. The latter authors also found that a relatively high ejection speed ($\sim100$~m/s) is required to form the stream if the ejection occurs at perihelion, a value that is too high to be explained by sublimation activity alone, but this can be easily compensated by the low precision of C/Mellish's currently-available orbit. The comet experienced close encounters with Venus in the past 1~kyr which further complicates knowledge of its past orbit. Linkages to two minor streams, the Daytime $\kappa$-Leonids and April $\rho$-Cygnids, have also been proposed \citep{Brown2008, Neslusan2014}. Comet C/Mellish will return around 2060. Recovery and observation of the comet will improve our knowledge of its orbit and physical characteristics, which will help understand its connection to these streams.

\subsection{LPCs}

Some long period comets that pass within 0.1~au from Earth's orbit cause occasional outbursts at intervals much shorter than the orbital period of the parents. It is now recognized that many of these outbursts are caused by the meteoroids ejected during the most recent perihelion passage of the parent \citep{Jenniskens1997c, Lyytinen2003}. Even so, details of when and how the meteoroids are ejected are still unclear. The existence of a few dozens of ``orphan'' LPC streams (streams without known parents) also needs to be understood.

\subsubsection{C/1861 G1 (Thatcher) and the April Lyrids}

Record of the April Lyrids (\#6) dates back to 1803 in which a strong outburst of ZHR=860 was recorded \citep{Jenniskens1995}. It may have been the meteor shower recorded by the Chinese observers in 687 BC \citep[][p. 11]{Jenniskens2006}, though without information of the radiant, it is difficult to verify this connection. The annual component of the shower peaks around April 22, reaching ZHR=20.

The shower appears to exhibit regular outbursts over a $\sim60$~yr interval, most recently in 1982, much shorter than the 415~yr period of C/Thatcher. Proposed explanations included a disrupted fragment in a 60~yr orbit \citep{Arter1995} or trapped dust trails in multiple resonances \citep{Emelyanenko2001}, however the ejection speed needed for these scenarios (on the order of several 100~m/s) is unrealistic based on the observed comet disruptions and sublimation models. Numerical modeling also showed that radiation drag could not bring optical-sized particles from parent orbits to the nearest plausible resonance within typical meteoroid collisional lifetime, or $\sim50$~kyr \citep{Kornos2015}. Alternatively, \citet{Jenniskens1997c} suggested that the outbursts from LPC streams can be explained by the reflex motion of the Sun due to Jupiter (and to a lesser extent, Saturn), a theory that has been verified by the successful prediction of the Aurigids outburst in 2007 (see below). The next April Lyrids outbursts in the cycle are in 2040/41.

\subsubsection{C/1911 N1 (Kiess) and the Aurigids}

The Aurigids (\#206) typically reaches ZHR=5 around September 1 every year. Several outbursts have been detected at irregular intervals, in 1935, 1986, 1994 and most recently in 2007. A connection to LPC C/1911 N1 (Kiess) was proposed shortly after the 1935 outburst was observed \citep[][\S~13.1]{Jenniskens2006}, but the fact that the Aurigid outbursts occurred in much shorter intervals compared to the $\sim2500$~yr orbital period of C/Kiess had been a puzzle for many years. \citet{Jenniskens1997c} introduced the reflex motion theory to explain the outbursts of LPC streams and predicted an Aurigid outburst in 2007 \citep{Jenniskens2007}. An outburst of ZHR=100 occurred as predicted \citep{Atreya2009}. 

The observed height distribution of the Aurigid meteors was in good agreement with the Leonids which has a similar entry speed, implying that the origin and composition of the Aurigids are likely compatible with that of the Leonids.

\subsubsection{C/1979 Y1 (Bradfield) and the July Pegasids}

The annual and weakly-active July Pegasids (\#175) has been detected in a number of video surveys but none of the radar survey, indicative of the dominance of larger, millimeter-class meteoroids. A connection to C/1979 Y1 (Bradfield) has been proposed by several authors and investigated by \citet{Hajdukova2019}. Numerical simulation has predicted encounters with other filaments which should produce meteor activity, but this has not been observationally confirmed.

\subsubsection{C/1983 H1 (IRAS--Araki--Alcock) and the $\eta$-Lyrids}

C/IRAS--Araki--Alcock made a close approach to the Earth at 0.03~au shortly after its discovery. Prediction of meteor activity was made, but no reliable detection of outbursts has been reported \citep[][\S~6.1]{Jenniskens2006}. A relatively strong annual $\eta$-Lyrid stream (\#145) has been detected in recent years, first from a few photographic orbits (\citealp{Jenniskens2006}, \S~6.1) and later from many low-light video orbits (\citealp{SonotaCo2009}; \citealp{Jenniskens2016c}).

\subsubsection{Other proposed linkages}

Besides the ones discussed above, there are about a dozen established meteoroid streams with proposed linkages to LPCs, as tabulated in Table~\ref{tbl:shr}. Most of these associations involve comets observed before modern astrometry becomes available, and thus it is difficult to critically examine the dynamical history of the comets (and therefore the linkages themselves). On the other hand, progress has been made in studying other LPCs and conducting a targeted search for the streams they may have produced \citep[e.g.][]{Hajdukova2019b, Hajdukova2020, Hajdukova2021, Jenniskens2021}. Most of the linkages being proposed involve streams that have not been confirmed. More comet and meteor data is needed to verify these proposed linkages.

\subsubsection{C/2013 A1 (Siding Spring) and a meteor storm at Mars}

C/Siding Spring passed Mars by a remarkable close distance of 0.0009~au on 2014 October 19, about 15 times closer than the close approach of D/1770 (Lexell) to the Earth. Despite the initial prediction that the bulk of the dust will miss Mars, significant enhancements of meteoric metal ions were detected by orbiters at Mars during the encounter \citep[cf.][\S~5.5.4]{Christou2019}. This may have pointed to distant activity of this LPC \citep[][\S~3.1]{Zhang2021b} which echoes the recent detection of the distant activity of LPCs like C/2017 K2 (PANSTARRS) \citep[cf.][]{Jewitt2021}.

\section{INTERRELATIONS}

\subsection{Physical and Dynamical Evolution of Comets}

Meteoroid streams trace the past evolution and activity of the parent comet, as far back as the coherence timescale of the meteoroid stream ($\sim10^3$--$10^5$~yr, \S~\ref{sec:comet2earth:demise}). Examples are the asteroidal meteoroid streams, which betray the fact that their parents must have been recently active. The Phaethon--Geminid Complex (\S~\ref{sec:obs:linkages:3200}) is an excellent example. Complexes that are simultaneously related to both comets and asteroids, such as the Machholz Interplanetary Complex (\S~\ref{sec:obs:linkages:96p}) and the 169P/NEAT Complex (\S~\ref{sec:obs:linkages:169p}), may indicate an inhomogeneous compositional structure of the progenitor which have been suggested from comet observations (see the chapter by {\it Guibert-Lepoutre et al.} in this volume). The low and sometimes the lack of meteor activity can be used to probe the recent history of low-activity comets and comet--asteroid transitional objects (see also the chapters by {\it Jewitt and Hsieh} and {\it Knight et al.} in this volume). Some comet examples were discussed in \S~\ref{sec:obs:linkages:low-activity}. By combining dynamical modeling, meteor data can provide additional insights into the activity of transitional objects. Investigation of transitional object 107P/(4015) Wilson--Harrington, for instance, has been limited by the singular activity detection in 1949. The absence of meteor activity of this Earth-approaching object (MOID=0.04~au) may imply a rather infrequent, or even a one-off, ejection. A cued search for meteor activity based on parent orbits revealed a meteor outburst from near-Earth asteroid (139359) 2001 ME$_1$ \citep{Ye2016c}, which may be similar to the activity of 107P in nature.

Similarly, significant meteor activity can be used to trace mass loss events, and even the demise, of the parent body. A large number of ``orphan'' meteoroid streams reveal frequent disruptions in the near-Earth space \citep{Vaubaillon2006}, and the relative abundance of near-Sun meteoroid streams can be explained by thermally disrupted asteroids \citep{Ye2019, Wiegert2020}. The Andromedid meteor storms (\S~\ref{sec:obs:linkages:3d}) would have revealed 3D's demise had the comet fragmentation not been observed. Meteor data and modeling of the $\alpha$-Capricornid and Phoenicid streams revealed previous fragmentations of their parents, 169P and 289P (\S~\ref{sec:obs:linkages:169p} and \S~\ref{sec:obs:linkages:289p}). Using meteor activity as a constraint to parent orbit was first demonstrated by \citet{Wiegert2013} on the case of 3D. Since meteor activity can often be traced to parent activity well before the telescopic era, this method can provide knowledge about the past orbit of the comet that is otherwise irretrievable. For example, application to the cold case of comet D/1770 L1 (Lexell), one of the largest near-Earth comets that is now lost, showed that the comet likely still remains in the solar system on an altered orbit \citep{Ye2018b}.

Finally, photometric, stereoscopic and spectroscopic characterization of meteors provide clues to the structure and composition of the materials that make up the parent. Mass photometry of meteors can be used to constrain the dust size distribution of the stream. Comparison with in-situ data from 1P/Halley and 81P/Wild 2 shows general agreement of the size distribution measured by spacecraft and by meteor observations (\citealp{Jenniskens2006}, \S~18.1; \citealp{TrigoRodriguez2021}). Meteor spectroscopy shows that most major streams have compositions broadly consistent with chondrites \citep{Borovicka2019}. Depletion of Na content is associated with near-Sun streams such as the Southern $\delta$-Aquariids from 96P/Machholz 1 and the Geminids from (3200) Phaethon \citep{Kasuga2006, Vojacek2015} due to thermal processing. A case-by-case discussions were given in \S~\ref{sec:obs:linkages}.

\subsection{Dust Trails and Meteoroid Streams}
\label{sec:inter:trails}

Meteoroid streams and optical/thermal dust trails detected by telescopes are essentially the same structure probed by different techniques. A case-by-case discussion has been given in \S~\ref{sec:obs:linkages}. Table~\ref{tbl:trails} summarizes the known dust trails originated from NEOs with MOID$<0.2$~au and their corresponding meteor showers (or the lack thereof). The lower limit of the total mass of a dust trail can be estimated using the brightness and dimension of the observed trail \citep[][\S~IV]{Sykes1992}. This lower limit tends to be overly conservative as the presumed mass of the typical dust in the trail tends to be underestimated, and that the trails often extend beyond the field of view. Compared to the total mass determined from meteor observations, we find that the determining mass of the dust trail is usually too low by 2--4 orders of magnitude. The mass of a meteoroid stream can be estimated using the method described by \citet{Hughes1989}. The largest source of uncertainty in this method is the mass influx, which has been better constrained thanks to the considerable progress in meteor surveys in the last few decades.

\begin{deluxetable}{cccccc}
\tablecaption{NEOs with significant (non-)detections of dust trails and/or meteor showers.\label{tbl:trails}}
\tablehead{
\colhead{Object} & \colhead{Optical?} & \colhead{IR?} & \colhead{Trail mass} & \colhead{Meteor shower} & \colhead{Stream mass} \\
} 
\startdata
1P/Halley & & \cmark $^{[1]}$ & $>4\times10^{7}$~kg & $\eta$-Aquariids, Orionids & $1-2\times10^{12}$~kg $^{[2]}$ \\
2P/Encke & \cmark $^{[3]}$ & \cmark $^{[1,4-7]}$ & $3-7\times10^{10}$~kg $^{[1,5]}$ & Taurids complex & $1\times10^{13}$~kg $^{[2]}$ \\
7P/Pons--Winnecke & & \cmark $^{[4]}$ & $>8\times10^8$~kg $^{[4]}$ & June Bootids & $6\times10^{8}$~kg/orbit \\
55P/Tempel--Tuttle & \cmark $^{[8]}$ & & $3\times10^{10}$~kg/orbit & Leonids & $5\times10^{12}$~kg $^{[9]}$ \\
73P/S--W 3 & & \cmark $^{[1,10]}$ & $>3\times10^6$~kg & $\tau$-Herculids & $5\times10^{8}$~kg \\
169P/NEAT & & \cmark $^{[1]}$ & $>3\times10^8$~kg & $\alpha$-Capricornids & $9\times10^{13}$~kg $^{[11]}$ \\
(3200) Phaethon & \cmark $^{[12]}$ & \cmark $^{[1]}$ & $>4\times10^{11}$~kg $^{[16]}$ & Geminids & $10^{13}$--$10^{15}$~kg $^{[13]}$ \\
46P/Wirtanen & \cmark $^{[14]}$ & \xmark & $>2\times10^5$~kg & - & $<5\times10^{7}$~kg \\
103P/Hartley 2 & & \cmark $^{[15]}$ & $4\times10^{10}$~kg $^{[15]}$ & - & n/a \\
107P/W--H & \xmark $^{[16]}$ & \xmark $^{[5]}$ & $\lesssim10^6$~kg? & - & $<10^{11}$~kg \\
P/2021 HS & \xmark $^{[17]}$ & & $<(3-5)\times10^{7}$~kg $^{[17]}$ & - & $<5\times10^{7}$~kg $^{[17]}$ \\
\enddata
\tablecomments{Total trail masses are either quoted from the references given, or derived using the measurements reported by the references following the method described by \citet{Sykes1992}. \cmark~in the optical and/or IR columns indicates detection of a dust trail in either/both wavelengths, while \xmark~indicates significant nondetection. References: [1] \citet{Arendt2014}; [2] \citet{Jenniskens1994}; [3] \citet{Ishiguro2007}; [4] \citet{Sykes1992}; [5] \citet{Reach2007}; [6] \citet{Kelley2006}; [7] \citet{Reach2000}; [8] \citet{Nakamura2000}; [9] \citet{Jenniskens2000b}; [10] \citet{Vaubaillon2010}; [11] \citet{Jenniskens2010}; [12] \citet{Battams2020}; [13] \citet{Ryabova2017}; [14] \citet{Farnham2019}; [15] \citet{Bauer2011}; [16] \citet{Ishiguro2011}; [17] \citet{Ye2022}}
\end{deluxetable}

\subsection{Interstellar Objects}

Studies of meteors of interstellar origin have been an important research topic in meteor science since the early 20th century. Despite extensive searches, no definitive detection has been made \citep[see][for a review]{Hajdukova2019}. On the other hand, successful detection of interstellar micrometeoroids has been made with dust detectors flown with a number of deep space missions \citep[e.g.][]{Westphal2014, Altobelli2016}.

The discovery of the first two interstellar objects, 1I/`Oumuamua and 2I/Borisov (see the chapter by {\it Fitzsimmons et al.} in this volume), the latter of which is an unambiguous comet, raises the possibility that meteoroids ejected from these objects can be detected. On its way in, the meteoroids released can sometimes disperse wide enough to be encountered by Earth. Such detection will allow materials from another planetary system to be sampled at Earth. Comet 2I/Borisov has $q=2.01$~au and does not come close to the Earth or any other major planets. 1I/`Oumuamua had an Earth MOID of $0.10$~au and a minimal fly-by distance of $0.16$~au, too far for meteoroid delivery driven by conventional water-ice sublimation. Indeed, a targeted search yielded no detection of meteor activity \citep{Ye2017}. The nature of 1I/`Oumuamua is debated, due to the fact that the object has an unusual disk-shape and exhibited non-gravitational acceleration with no coma \citep{Micheli2018}, and it can only be agreed that the object is neither a typical asteroid nor a typical comet. A number of hypotheses have been put forward to explain `Oumuamua's erratic motion, including the sublimation of molecular hydrogen or nitrogen ice \citep{Hoang2020, Seligman2020, Desch2021}. `Oumuamua's distant fly-by only permits the H$_2$ ice scenario to be tested using meteor observation, and only a loose constraint of $<10~\mathrm{kg~s^{-1}}$ was derived \citep{Ye2017} which was several orders of magnitude larger than sustainable considering `Oumuamua's small mass ($\sim10^9$~kg). However, future sensitive meteor observations (perhaps through dedicated campaigns if advance warning can be given) could permit more stringent limits to validate these models.

\subsection{Sample Return as (Micro-)Meteorites}

Meteoroids larger than a few 0.1~m may survive the atmospheric entry and reach the ground as meteorites. Thousands of meteorites have been found on Earth, but only a handful of them (38 at the time of writing) have been observed instrumentally for which orbits of the pre-entry body can be derived, and none have been unambiguously associated with comets. The expected mineral and chemical properties of cometary meteorites are under debate. If analogous to cometary dust, they would likely be nearly chondritic with a high abundance of C, N, and organics \citep{Campins1998}. It has long been suspected that CI chondrites may have come from comets \citep[cf.][]{Lodders1999}, but this scenario appears increasingly unlikely given that {\it Hayabusa2} samples from (162173) Ryugu, an undisputed asteroid, are consistent with CI chondrites. Recent comet missions such as {\it Rosetta} and {\it Stardust} have measured cometary dust and revealed some similarities to primitive meteorites \citep{Stadermann2008, Hoppe2018}, but do not offer substantially more information for the search of cometary meteorites. Meanwhile {\it Philae}, designed to land on 67P/Churyumov--Gerasimenko and directly measure surface and subsurface material, only returned limited data. The derived properties are difficult to reconcile with other observations and are hence inconclusive \citep{Keller2020}.

Dynamical properties (i.e. orbits) provide an easy but somewhat ambiguous metric to identify possibly cometary meteorites. Based on the orbit derived from unaided witness reports, it has been proposed that the historic Orgueil meteorite may be cometary \citep{Gounelle2006} but orbits derived this way can be highly uncertain \citep{Egal2018b}. The 1908 Tunguska event \citep{Jenniskens2019b} has also been associated with a comet, specifically a fragment of 2P/Encke \citep[cf.][]{Kelley2009}, but that too remains controversial. Analysis of meter-sized impactors showed that 10--15\% of them have comet-like orbits \citep{Brown2016}. Compared to asteroids, comets tend to have higher relative velocities with respect to the Earth and lower mechanical strength, making any meter-class cometesimals harder to survive  atmospheric penetration. Specifically, it has been found that Taurid meteors brighter than -7 magnitude are in the strength-limited regime and do not penetrate deeper with increasing size \citep{Devillepoix2021}.

At the other end of the spectrum, micrometeoroids (meteoroids smaller than $\sim200~\micron$) are too small to ablate during atmospheric entry, and hence can survive to the ground. They are mostly found on ocean floors and in polar sediments where the surrounding environment is largely stable, though fresh micrometeoroids can be collected from an urban environment with careful treatment \citep{Genge2017, Suttle2021}. Unlike meteorites, the vast majority of micrometeoroids are compositionally compatible with carbonaceous chondrites, indicative of cometary origin \citep{Engrand1998, Nesvorny2010}.

\subsection{Astrobiology}

Organic matter and water were delivered to Earth during or before the late heavy bombardment about 4~Gyr ago, creating the conditions possible for life. The mass influx of meteoroids is of the same order as that of planetesimals \citep{Jenniskens2004f}. Meteors also deliver organics and metalic compounds that can have fertilized oceans and supplied organic compounds. The metallic content in meteoroids can be probed via atomic and ionic emissions of meteors, of which many are accessible in the visible and near-infrared wavelengths. Elements such as Ng, Mg, Fe, Ca and Si can easily be explored using visible spectroscopy \citep[cf.][]{Borovicka2019, Koten2019}. Organic matter produces CN bands, hydrogen and OH emissions. Attempts have been made to look for the strong CN emission at 387~nm and the OH (0-0) emission at 308~nm but without clear detection \citep{Jenniskens2004g, Abe2007}. \citet{Jenniskens2004f} derived an upper limit of $<0.017$ for CN/Fe, more than two orders of magnitude lower than if all nitrogen from a cometary dust is converted to CN. Simulation using a thermochemical equilibrium model suggested that most of the carbon atoms from organic matter in plasma excitation temperature around 4000--5000~K quickly bound with atmospheric oxygen and form CO, and few form CN and C$_2$, making detection even more difficult \citep{Berezhnoy2010}. On the other hand, thermochemical equilibrium is not expected during meteoroid entry. Regarding OH, the same model also showed that it is possible to satisfy several constraints derived by \citet{Jenniskens2004f, Abe2007} under certain temperature ranges \citep[see][\S~6]{Berezhnoy2010}. However, the presence of the OH (0-0) emission needs further confirmation. It is also not clear how much of the plausible OH feature is produced by the reaction by the ozone--hydrogen reaction rather than the dissociation of meteoric mineral water. The latter scenario, if confirmed, can provide another way to provide hydrated content delivered to the Earth from comets and water-rich asteroids.

\subsection{Satellite Impact Hazard}
\label{sec:inter:sat}

Meteoroid impacts can interfere with, and sometimes terminate, the operation of satellites and spacecraft in orbit. Adverse effects from meteoroid impacts include mechanical damage to the hardware and unwanted electronic effects caused by impact-generated plasma, which can result in life-threatening situations for astronauts. Hazards from meteoroid impacts have received great attention in recent years, and various preventive measures are being taken to mitigate the risk of an impact, including the development of meteoroid models, improved spacecraft designs, and operational measures such as changing the orientation of solar panels or performing maneuvers to minimize the surface area exposed to a shower. Cometary streams, especially the ones from HTCs and LPCs, pose a greater threat due to their higher arrival speed at Earth. This is one of the motivations to the study of such meteor showers and outbursts. Detailed reviews of this topic include \citet{Drolshagen2019}.

\subsection{Planetary Defense}

Meteoroid streams can also betray undiscovered parents which are of significance in planetary defense. After several decades of NEO surveying, $>95\% $km-class NEAs have been cataloged \citep{Jedicke2015}, but km-class LPCs are much less known and now are the dominant remaining impact risk in planetary defense. There are presently several dozens of established HTC/LPC meteoroid streams that have not been identified with any known comets or asteroids, and more await confirmation. The strongest and the best-characterized among them is the $\alpha$-Monocerotids (\#246), an annual meteor shower with activity in mid-November which has also produced handful intense outbursts \citep{Jenniskens1997, Jenniskens1997b}. All LPC showers with such outbursts have parent bodies that oscillate in a similar manner in and out of Earth orbit and thus are a potential impact hazard.

\section{FUTURE WORK}

Since {\it Comets II}, our understanding of comets and their meteoroid streams has advanced significantly. This trend is expected to continue given the development on many fronts.

On the front of meteor science and observation, more optical and radar meteor networks are entering operation, especially over areas that have had poor coverage in the past. Networks such as CAMS, the Global Fireball Observatory, Fireball Recovery and InterPlanetary Observation Network (FRIPON), and the Global Meteor Network are expanding globally, and new video networks and meteor radars are being set up in historically poorly-covered regions. The accumulation of data allows existing streams to be better characterized and new streams to be verified. Advances in sensor technology pave the way for instruments with higher sensitivity, frame rate and/or resolution, enabling better constraints on meteoroid properties and orbits.

On the front of traditional telescopic science, the next-generation astrophysical observatories and time-domain surveys, such as Vera Rubin Observatory/Legacy Survey of Space and Time (LSST), NEO Surveyor, James Webb (JWST), Roman, Euclid, and the Chinese Space Station Telescope (CSST), will enhance our understanding of various small body populations (see the chapter by {\it Bauer et al.} in this volume), resulting likely in the detection of more meteoroid stream parents or remnants from past breakup events. LSST alone will increase the number of cataloged small bodies by 5--10$\times$ \citep{Schwamb2018}. In addition, LSST and numerous shallower time-domain surveys will continue to monitor episodic ejection events from comets that are responsible for the formation of some streams in the near future. Other multi-bandpass and wavelength surveys (e.g. NEO Surveyor, Roman, Euclid, CSST), as well as multi-usage telescopes (e.g. JWST), will characterize the parents and provide context for meteor observations.

On the front of in-situ exploration, new small-body missions will provide insights into the origin and evolution of the parent bodies as well as the formation of the meteoroid streams. The {\it Rosetta} mission has, along with many other things, provided crucial data for understanding the formation of meteoroid stream at the source, even though 67P/Churyumov–-Gerasimenko does not currently produce a meteor shower at Earth. In the foreseeable future, we have at least {\it Comet Interceptor} that will visit a to-be-discovered long-period comet or interstellar comet, with a handful of backup targets that include many meteor shower parents, as well as {\it DESTINY$^+$} to Phaethon and 2005 UD. Knowledge from meteor observation/stream modeling will complement the data returned from these missions.

Other Earth-based and interplanetary missions can potentially provide unique data for meteor research as well. Weather radars and satellites occasionally assist investigation of bolides \citep[e.g. GLM,][]{Jenniskens2018c}. The unexpected flyby of C/2013 A1 (Siding Spring) to Mars allowed the spacecraft at Mars to observe the comet as well as the meteor shower in Mars atmosphere up close, providing a wealth of data of this dynamically new comet. The event is a warning that more intense meteor showers are possible than we have experienced in recent years. A more distant flyby of C/2021 A1 (Leonard) to Venus at the end of 2021 provided an opportunity to examine the activity of the comet at $>30$~au \citep{Zhang2021b}, made possible with the Akatsuki spacecraft which is currently orbiting the planet. Mars is sparsely populated with landers and rovers, which can conduct complimentary meteor observations. Other planets and moons with notable atmospheres or exospheres are also ideal places to study meteors, such as Mercury's exosphere and impact flashes on Moon and Jupiter. These data provide additional points to sample and study dust streams in the solar system in addition to our Earth. With the ever-growing interest to explore our solar system with in-situ missions, the opportunities to study these ``exo'' meteors are numerous.

We conclude with some questions for future research:

\begin{enumerate}
    \item What is the formation mechanism of asteroidal meteoroid streams? How to use meteor data to distinguish dormant comets from asteroids?
    \item What is the underlying cause of the streams with simultaneous linkages to asteroids and comets? Are these linkages coincidental?
    \item How to use meteor data to identify and investigate dormant comets in the NEO population? What is the mechanism that drives episodic ejections of 107P/Wilson--Harrington and (139359) 2001 ME$_1$?
    \item How to use meteor data to search for hidden LPC impactors? Can we use meteor data to probe the ``fading paradox'' of the LPCs?
    \item What is the cause of a large number of orphan meteoroid streams, especially towards the HTC/LPC regime? Is this due to the observational bias of the telescopic surveys?
    \item How to effectively distinguish real parent--stream linkages from chance alignments in the big data era? 
\end{enumerate}

\noindent \textbf{Acknowledgments.} \\
We thank Marc Fries, Jeremie Vaubaillon and an anonymous reviewer for their careful review, as well as David Jewitt, Summer Xia Han, and Matthew Knight for their comments, all of which help improve this chapter. We also thank Karl Battams, Zhuoxiao Wang and Steed Yu for kindly providing their images of the Geminids. QY was partially supported by NASA grant 80NSSC22K0772. PJ was supported by NASA grant 80NSSC19K0563.

\bibliographystyle{sss-three.bst}
\bibliography{refs.bib}

\end{document}